\documentclass[10pt,aps,prd,onecolumn,amsmath,amssymb,superscriptaddress,nofootinbib]{revtex4}

\usepackage{graphicx}
\usepackage{subcaption}
\usepackage{epsfig}
\usepackage{bm}
\usepackage[normalem]{ulem}
\usepackage{color}
\usepackage{multirow}
\usepackage{slashed}

\usepackage{hyperref}
\usepackage{verbatim}
\usepackage{url}
\usepackage{braket}
\usepackage{amssymb}
\usepackage{adjustbox}

\usepackage{tikz}
\usetikzlibrary{arrows,shapes}
\usetikzlibrary{trees}
\usetikzlibrary{matrix,arrows} 				
\usetikzlibrary{positioning}				
\usetikzlibrary{calc,through}				
\usetikzlibrary{decorations.pathreplacing}  
\usepackage{pgffor}							

\usetikzlibrary{decorations.pathmorphing}	
\usetikzlibrary{decorations.markings}
\tikzset{
	vector/.style={decorate, decoration={snake}, draw},
	provector/.style={decorate, decoration={snake,amplitude=2.5pt}, draw},
	antivector/.style={decorate, decoration={snake,amplitude=-2.5pt}, draw},
	fermion/.style={draw=black, postaction={decorate},
		decoration={markings,mark=at position .55 with {\arrow[draw=black]{>}}}},
	fermiona/.style={draw=red},
	fermionbar/.style={draw=black, postaction={decorate},
		decoration={markings,mark=at position .55 with {\arrow[draw=black]{<}}}},
	fermionnoarrow/.style={draw=black},
	gluon/.style={decorate, draw=black,
		decoration={coil,amplitude=4pt, segment length=5pt}},
	scalar/.style={dashed,draw=black, postaction={decorate},
		decoration={markings,mark=at position .55 with {\arrow[draw=black]{>}}}},
	scalarbar/.style={dashed,draw=black, postaction={decorate},
		decoration={markings,mark=at position .55 with {\arrow[draw=black]{<}}}},
	scalarnoarrow/.style={dashed,draw=black},
	electron/.style={draw=black, postaction={decorate},
		decoration={markings,mark=at position .55 with {\arrow[draw=black]{>}}}},
	bigvector/.style={decorate, decoration={snake,amplitude=4pt}, draw},
}

\usepackage{hyperref}

\def\ben{\begin{equation}}
	\def\een{\end{equation}}

\def\be{\begin{equation}}
	\def\ee{\end{equation}}
\def\beq{\begin{equation}}
	\def\eeq{\end{equation}}
\def\ba{\begin{array}}
	\def\ea{\end{array}}

\def\dalemb#1#2{{\vbox{\hrule height .#2pt
			\hbox{\vrule width.#2pt height#1pt \kern#1pt
				\vrule width.#2pt}
			\hrule height.#2pt}}}

\newcommand{\bea}{\begin{eqnarray}}
	\newcommand{\eea}{\end{eqnarray}}

\newcommand{\trace}{\mbox{Tr}}

\newcommand{\red}[1]{\textcolor{red}{#1}}

\DeclareDocumentCommand{\nint}{ O{} O{} m }{\ensuremath{ \int_{\mbox{\scriptsize $#1$}}^{\mbox{\scriptsize$#2$}}\!\!\! \mbox{\small $\,\mathrm{d}#3$\! }}}

\begin{document}
	\title{Excitation Transmission through a non-Hermitian traversable wormhole}
	\author{Sizheng Cao}
	\affiliation{Department of Physics, College of Sciences, Shanghai University, 99 Shangda Road,
200444 Shanghai, China}
	\author{Xian-Hui Ge~\footnote{ Corresponding author. gexh@shu.edu.cn}}
	\affiliation{Department of Physics, College of Sciences, Shanghai University, 99 Shangda Road,
200444 Shanghai, China}
\affiliation{Institute for Quantum Science and Technology,  Shanghai University, 99 Shangda Road,
200444 Shanghai, China}


	\date{\today}
	\begin{abstract}
		This study explores the intricate real-time dynamics of a non-Hermitian system composed of two interconnected Sachdev-Ye-Kitaev (SYK) models. A central finding reveals that an excitation initially localized in the right SYK subsystem can be efficiently transmitted to the left subsystem subsequent to the characteristic scrambling time, a phenomenon facilitated by the intrinsic non-Hermitian nature of the system. The defining hallmark of non-Hermiticity is manifest in the asymmetric conveyance of quantum states, with the non-Hermitian parameter functioning as a tunable knob that selectively amplifies or dampens propagation modes on either side. Despite this inherent directional bias in state transfer, the system sustains two distinct phases, analogously likened to black holes and wormholes.
	\end{abstract}

	\maketitle

\section{Introduction}
In the landscape of modern physics, non-Hermitian Hamiltonians have emerged as a fecund and thought-provoking domain of inquiry, transcending the conventional boundaries of Hermitian systems and offering a wealth of unconventional phenomena and novel applications \cite{Bender_2007,Arean_2020,Bender_1998,Bender_2005,Bender_2005_2,PhysRevLett.124.196401,PhysRevLett.118.045701,PhysRevB.97.041203,PhysRevB.97.075128,PhysRevLett.77.570,He_2022,nosaka2020quantum,S_nderhauf_2019}. Their unique attributes, such as exceptional points \cite{PhysRevLett.112.203901,PhysRevA.93.033809,Berry2004,Heiss_2004,PhysRevB.97.121401}, skin effects \cite{Yao:2018eyg,PhysRevB.99.201103,Li2020}, and parity-time symmetry breaking \cite{garcíagarcía2023classification,Weimann2017,PhysRevB.97.201411,Wu_2019}, have found relevance in diverse fields ranging from optics and quantum simulators to topological materials and sensing technologies \cite{El-Ganainy2019,PhysRevLett.106.213901,Rüter2010,El-Ganainy2018,Suchkov_2016,feng_sci}. Concurrently, the interface between Hermitian/non-Hermitian physics and holography has garnered considerable interest, representing a bold attempt to bridge the gap between strongly correlated quantum matter and higher-dimensional gravitational theories. A central player in this exploration is the SYK model \cite{1604-07818,1804-00491,PhysRevD.100.026007,susskind2022scrambling,PhysRevX.8.021049,PhysRevResearch.2.043049,Peng_2021,Peng_2017,PhysRevD.108.086014,cao2021thermodynamic,Cai2018,PhysRevD.106.046008,softmode}, a minimal yet powerful exemplar of quantum many-body chaos, which owes its significance to its exact holographic duality with a two-dimensional Jackiw-Teitelboim gravity theory. This duality provides a direct correspondence between the intricate dynamics of the SYK model and the geometric and gravitational attributes of its holographic counterpart, thereby serving as a fertile ground for examining the interplay of strongly coupled quantum elements within the holographic framework.

One particularly captivating extension of the SYK model is the realm of non-Hermitian two-coupled SYK systems. By endowing the original two coupled SYK model--the Maldacena-Qi model \cite{1804-00491} with non-Hermitian couplings, researchers have ventured into largely uncharted territory where the interplay between non-unitary dynamics, quantum correlations, and holographic correspondence can yield groundbreaking insights. Initial inquiries, as chronicled in \cite{2208-10800} by some of the authors, have already shed light on the resilience of the thermodynamic structure in the face of non-Hermitian perturbations, suggesting that certain core aspects of the SYK model remain impervious to the introduction of non-conservative interactions.

In this work, we focus specifically on the revival dynamics of the non-Hermitian two coupled SYK system, seeking to unravel a strikingly unique and insightful physical narrative that transcends prior understandings. Our investigation delves into the intricate interplay between the non-Hermitian couplings and the emergent collective behavior of the coupled SYK subsystems, which, in the holographic context, may correspond to novel gravitational or topological features in the dual spacetime.  This non-Hermitian dual-SYK paradigm delineates a physics scenario wherein, in the absence of coupling ($\mu=0$)
, the Hamiltonian essentially decomposes into two standalone Hermitian SYK models. Consequently, when initialized in the ground state with an excitation introduced on the right subsystem via $ C^{\dagger}_R|\psi_0\rangle
$ , the quantum excitation rapidly diffuses and scrambles across an immense multitude of states within the Hilbert space of the right-side SYK Hamiltonian, 
$H^R_{\rm SYK}$, rendering it virtually untraceable by conventional observables. 
Upon introducing a weak  coupling 
$\mu$ plus non-Hermitian terms, however, a transformative dynamic ensues. The scrambled excitation undergoes a directional transfer to the left side within a characteristic timescale 
$t_{re}$ . Subsequently, the system evolves to a state akin to  $ C^{\dagger}_L|\psi_0\rangle$  . The essence of non-Hermiticity manifests in the asymmetry of quantum state transmission; the non-Hermitian parameter assumes a pivotal role, acting as a tunable amplifier or suppressor for the propagation modes on either the left or right sides. This phenomenon profoundly impacts the directional transport of quantum excitation, offering a novel perspective on the intricate interplay between non-unitary dynamics and scrambling in coupled SYK systems. In this paper, we perform a comprehensive numerical study of several specific quantities to support our results. We firstly investigate the non-diagonal real-time elements of the Green's functions, \(G_{LR}\) and \(G_{RL}\) for black hole and wormhole phases, as well as their corresponding spectral functions. Additionally, we analyze the diagonal elements of the Green's functions and their respective spectral functions to understand the system's behavior more thoroughly. To provide further insights, we examine the imaginary parts of the Green's functions and observe the influence of the non-Hermitian parameter $\alpha$ on the frequency of the first peak of the spectral functions. We also calculate the transmission amplitude of the Green's functions, which reflects the tunneling amplitude between different sites. Moreover, we explore the non-diagonal elements of the Green's functions at finite $N$ and study the out-of-time-ordered correlators (OTOCs) under different values of $\alpha$ and temperature conditions. These quantities collectively offer a detailed understanding of the system's dynamics and the impact of non-Hermiticity.
This addition provides a clearer explanation of the specific quantities we study and how they support the results claimed in our paper.

The structure of this paper is outlined as follows: Section II commences with the introduction of the non-Hermitian Hamiltonian as a pivotal component of our study. Subsequently, in the same section, we delve into an in-depth analysis of particle transmission through an asymmetrically traversable wormhole configuration. Progressing to Section V, we scrutinize the behavior of OTOCs, which play a significant role in characterizing the chaotic dynamics of the system. Finally, we conclude with a synthesis of our findings and engage in a comprehensive discussion in the concluding section.

\section{The non-Hermitian model}\label{model}
We start with the Hamiltonian describing two-coupled non-Hermitian SYK model. The Hamiltonian reads
\begin{align}\label{hamilton}
    &H=H^L_{\rm SYK}+H^R_{\rm SYK}+H_{int},\nonumber\\
    &H_0=H^L_{\rm SYK}+H^R_{\rm SYK}=-\sum_{i,j,k,l=1}^N J_{ijkl}\sum_{A=L,R}\left( 2
C_i^{A\dagger}C_j^{A\dagger}C_k^{A}C_l^{A}+4C_i^{A\dagger}C_j^{A}C_k^{A\dagger}C_l^{A}\right),\nonumber\\
&H_{int}=i\mu\sum_{i=1}^N(e^{-2\alpha}C_i^{L\dagger}C_i^{R}-e^{2\alpha}C_i^{R\dagger}C_i^{L}),
\end{align}
where  $\mu$ is the coupling strength (not the chemical potential \footnote{We work in the zero chemical potential condition, which corresponds to particle-hole symmetric case.}) that connect the two sides L and R, and the parameter $\alpha$ controls the non-Hermiticity of our model.  
$C_i$ and $C_i^\dagger$ are Dirac fermion operators satisfying the following anti-commutation relations
\begin{align}
\{C_i^A, C_j^{B\dagger}\} = \delta_{ij}\delta_{AB}, \quad \{C_i, C_j\} = \{C_i^\dagger, C_j^\dagger\} = 0.
\end{align}
These operators substitute the Majorana fermions used in the traditional Majorana SYK model and facilitate the introduction of non-Hermiticity through a similarity transformation, as detailed in the appendix A of \cite{2208-10800}.

This non-Hermitian two coupled SYK model can provide thermodynamic structure equivalent to a Hermitian two coupled SYK model \cite{2208-10800}. The findings indicate that the non-Hermiticity does not exert any influence on the energy spectrum, entanglement degree of the ground states, or the low-energy effective action associated with this model. This is attributed to the fact that the Hamiltonian provided in eq.\eqref{hamilton}, which is essentially a pseudo-Hermitian Hamiltonian, can be derived through a self-similarity transformation from the MQ model while eliminating unphysical terms. The eigen-energies of the Hamiltonian are always real within the parameter range $\alpha\in \mathbb{C}$. 
In a more precise manner, we can utilize the exact diagonalization method to numerically construct and solve the 
$2^{2N}\times2^{2N}$ 
  dimensional Hamiltonian matrix representative of a finite-sized (
$N$ fermions) system. Within a single computational realization, we deliberately fix the set of Gaussian-distributed random couplings  $J_{ijkl}$
 to ensure their constancy. This controlled setup enables us to observe that distinct values of the non-Hermiticity parameter $\alpha$
 and  $\alpha'$  can yield an entirely identical spectrum of eigen-energies, demonstrating a surprising insensitivity of certain eigenstates to variations in the degree of non-Hermiticity. The novel biorthogonal ground states reveal that, by adjusting the non-Hermiticity, two SYK sites can be independently tuned; one can occupy the ground state, while the other transitions to the Schwarzian excited state. Notably, the free energy remains unaffected by the non-Hermiticity, maintaining the same thermodynamic structure. However, we anticipate that the dynamical properties of this non-Hermitian Hamiltonian will exhibit distinctive characteristics, as the real-time evolution of the Green function strongly depends on the non-Hermiticity. This suggests that the non-Hermitian Hamiltonian may offer novel insights into the dynamic behavior of the system, complementing the static properties captured by the eigen-energies and ground states.

As we have discussed above, all thermodynamic properties remain unaffected by non-Hermitian properties. This prompts us to analyze the non-Hermitian model from an alternative perspective. Essentially, the model is derived from a similarity transformation of a Hermitian Hamiltonian. Furthermore, any two distinct values of the non-Hermitian parameter $\alpha$ share an identical set of energy eigenvalues. This can explains why the partition function and free energy remain invariant between Hermitian and non-Hermitian models. In general, the Hamiltonian yields the transformation by a matrix $\mathcal{S}$  
\begin{align}
H^\prime(\alpha^\prime)=\mathcal{S}^{-1} H(\alpha) \mathcal{S},
\end{align}
here we label the $\alpha$-denpendent Hamiltonian as $H(\alpha)$. Notice that the exponential of a matrix satisfies the relation $e^{\mathcal{S}^{-1}A\mathcal{S}}=\mathcal{S}^{-1}e^A\mathcal{S}$. From the definition of partition function we have
\begin{align}    Z^\prime[H^\prime(\alpha^\prime)]=\trace \left[e^{-\beta \mathcal{S}^{-1}H(\alpha) \mathcal{S}}\right]=\trace\left[\mathcal{S}^{-1}e^{-\beta H(\alpha) }\mathcal{S}\right]=\trace\left[e^{-\beta H(\alpha) }\right]=Z[H(\alpha)].
\end{align}
 However, when considering the two-point correlator, the introduction of additional operators $\mathcal{S}$ complicates the situation.  Taking the greater Green's function as an illustrative example,
\begin{align}\label{Green_Function}
    G_{AB}^>(t_1,t_2;\alpha)=\sum_{i}\frac{\theta(t)}{N}\braket{C_i^A(t_1)C_i^{B\dagger}(t_2)}_\beta,
\end{align}
in which the $\braket{\dots}_\beta$ represents thermal ensemble average $\trace\left[e^{-\beta H}\dots\right]$, for simplicity we set the $t_1=t>0$ and $t_2=0$ and leave out the greater symbol in the following content i.e. $G^>_{AB}(t_1,t_2;\alpha)=G_{AB}(t;\alpha)$.
The time evolution of any operator in the real time is defined as
\begin{align}
\mathcal{O}(t)=e^{iHt}\mathcal{O}(0)e^{-iHt},
\end{align}
with this and \eqref{Green_Function} we have
\begin{align}
    G_{AB}(t;\alpha^\prime)&=\frac{1}{N}\sum_i\trace\left[
    \mathcal{S}^{-1}e^{-\beta H}\mathcal{S} \mathcal{S}^{-1}e^{iHt}\mathcal{S}C_i^A(0)\mathcal{S}^{-1}e^{-iHt}\mathcal{S}C_i^{B\dagger}(0)\right]\nonumber\\
    &=\frac{1}{N}\sum_i\trace\left[
    e^{-\beta H}e^{iHt}(\mathcal{S}C_i^A(0)\mathcal{S}^{-1})e^{-iHt}(\mathcal{S}C_i^{B\dagger}(0)\mathcal{S}^{-1})
    \right].
\end{align}
Unlike the Hamiltonian, which remains invariant under certain transformations, the correlators in non-Hermitian models exhibit a pronounced dependence on the similarity transformation matrix, $\mathcal{S}$.  Essentially, the analysis presented above holds universally for the thermodynamic states of arbitrary quantum systems. This implies that two systems that are related by similarity transformations share identical thermodynamic properties, yet they can be dynamically distinguished from each other. 


The Euclidean time Schwinger-Dyson equations of the model can be analytically extended to the Lorentzian time domain via the Wick rotation, where $\tau$ is replaced by $it$ and $\omega$ is replaced by $i\omega_n$. This transformation leads to a set of corresponding real-time SD equations
\begin{align}\label{SDequation}
&\Sigma^>_{AB}(t)=-36J^2G^{>2}_{AB}(t
)G^>_{BA}(-t
),\nonumber\\
&\tilde{G}^R_{LL}(\omega,\alpha)=\frac{-\omega-\tilde{\Sigma
}^R_{RR}(\omega,\alpha)}{\tilde{D}(\omega,\alpha)},\nonumber\\
&\tilde{G}^R_{RR}(\omega,\alpha)=\frac{-\omega-\tilde{\Sigma
}^R_{LL}(\omega,\alpha)}{\tilde{D}(\omega,\alpha)},\nonumber\\
&\tilde{G}^R_{LR}(\omega,\alpha)=\frac{-i\mu e^{-2\alpha}+\tilde{\Sigma
}^R_{LR}(\omega,\alpha)}{\tilde{D}(\omega,\alpha)},\nonumber\\
&\tilde{G}^R_{RL}(\omega,\alpha)=\frac{i\mu e^{2\alpha}+\tilde{\Sigma
}^R_{RL}(\omega,\alpha)}{\tilde{D}(\omega,\alpha)},\nonumber\\
&\tilde{D}(\omega,\alpha)=\left( -\omega-\tilde{\Sigma
}^R_{LL} \right)\left( -\omega-\tilde{\Sigma
}^R_{RR} \right)+\left( i\mu e^{-2\alpha}-\tilde{\Sigma
}^R_{LR} \right)\left( i\mu e^{2\alpha}-\tilde{\Sigma
}^R_{RL} \right).
\end{align}
To enhance clarity, we introduce the notation $t=t_1-t_2$ to streamline the equations. Throughout this work, $\omega$ represents Euclidean time, while $\omega_n$ denotes the Matsubara frequency. Additionally, the labels $t$ and $\omega$ are used to signify Lorentzian real time and frequency, respectively. See appendix \ref{appendix1} for detailed derivation of the SD equations. To maintain brevity, we will omit the greater sign in the greater Green's functions, thus utilizing the notation  $G^>_{AB}=G_{AB}$ in subsequent sections. Typically, in the presence of mirror symmetry, the SD equations are simplified to two pairs, namely $LL$ and $LR$. However, the introduction of non-Hermitian models disrupts this symmetry, necessitating the consideration of the complete set of SD equations. 

The term `mirror symmetry' mentioned 
in \cite{1804-00491,PhysRevResearch.2.043049,2003-03914} refers to the specific transformation \( C_i^L \to C_i^R \) and \( C_i^R \to -C_i^L \). In the Hermitian case where \(\alpha = 0\), the Hamiltonian remains invariant under this transformation and the transformation send $G_{LR}\rightarrow-G_{RL}$. Whereas the physical processes described by the Green's functions remain unchanged under the transformation. Consequently, the Green's functions \(G_{LL}\) and \(G_{RR}\) are equal, and the off-diagonal Green's functions satisfy \(G_{LR} = -G_{RL}\). This implies
\[ G_{LL}(t) = G_{RR}(t) ~,\quad G_{LR}(t) = -G_{RL}(t) ~.\]
However, when \(\alpha\) is non-zero, the Hamiltonian changes under the same transformation, leading to the breaking of this mirror symmetry. Specifically, the introduction of the non-Hermitian parameter \(\alpha\) results in
\( G_{LR}(t) \neq -G_{RL}(t) \). This breaking of symmetry is a crucial aspect of our study, as it directly influences the directional transmission of quantum excitation within the system. Therefore, the non-Hermitian parameter \(\alpha\) introduces additional complexities into the system's dynamics. This comprehensive approach adds computational complexity to numerical calculations, particularly in the low-temperature regime, thereby posing greater challenges in obtaining accurate solutions. This comprehensive approach adds computational complexity to numerical calculations, particularly in the low-temperature regime, thereby posing greater challenges in obtaining accurate solutions.

The retarded two point functions and self-energies are defined as   

\begin{equation}
    G^R_{AB}(t)=\theta(t)(G_{AB}^>(t)-G_{AB}^>(-t)),\quad
    \Sigma_{AB}^R=\theta(t)(\Sigma_{AB}^>(t)-\Sigma_{BA}^>(-t)),
\end{equation}
where $\theta(t)$ is the Heaviside step function. 
In frequency space $\omega$, the greater two point function $\tilde{G}^>_{AB}(\omega)$ and the spectral function $\tilde{G}^R_{AB}(\omega)$ obey the relation \cite{Mahan1990}
\begin{align}\label{greater}
    &\tilde{G}^>_{LL}(\omega)=\frac{i\rho_{LL}(\omega)}{1+e^{-\beta\omega}},\quad \tilde{G}^>_{LR}(\omega)=\frac{\rho_{LR}(\omega)}{1+e^{-\beta\omega}},\nonumber\\
    &\tilde{G}^>_{RR}(\omega)=\frac{i\rho_{RR}(\omega)}{1+e^{-\beta\omega}},\quad \tilde{G}^>_{RL}(\omega)=\frac{\rho_{RL}(\omega)}{1+e^{-\beta\omega}},
\end{align}
by introducing the spectral functions $\rho_{AB}(\omega)$ are 
\begin{align}\label{spectral}
    &\rho_{LL}(\omega)=-2{\rm Im}[\tilde{G}^R_{LL}(\omega)],\quad \rho_{LR}(\omega)=-2{\rm Re}[\tilde{G}^R_{LR}(\omega)],\nonumber\\ &\rho_{RR}(\omega)=-2{\rm Im}[\tilde{G}^R_{RR}(\omega)],\quad \rho_{RL}(\omega)=-2{\rm Re}[\tilde{G}^R_{RL}(\omega)].
\end{align}
The equations\eqref{SDequation},\eqref{greater} and \eqref{spectral}  constitute a set of self-consistent equations. We can employ the same iterative algorithm to numerically solve the SD equation. Previous research findings indicate that the non-Hermitian interaction does not lead to complex values for the energy spectrum and thermodynamic quantities. In other words, the system energy remains entirely real, independent of the non-Hermitian parameter $\alpha$. 
However, upon examining the SD equation~\eqref{SDequation}, it becomes evident that the left-right and right-left correlators, $G_{LR}$
  and  $G_{RL}~$, explicitly depend on the non-Hermitian parameter $\alpha$. This dependence indicates that the mirror symmetry or $Z_4$
  symmetry has been broken, resulting in the inequality $G_{LR}(t)\neq -G_{RL}(t)$.  Actually, the propagators of the SD equation satisfy the specific relation below
\begin{align}\label{new_mirror_symmetry}
    G_{LL}(t;\alpha)=G_{RR}(t;\alpha),\quad e^{2\alpha}G_{LR}(t;\alpha)=-e^{-2\alpha}G_{RL}(t;\alpha),
\end{align}
when $\alpha=0$, the model return to the Hermitian MQ's model and the mirror symmetry recovers.  Intriguingly, there are two different symmetries for the Hamiltonians: $H_0$ is $\alpha$-independent and satisfies the symmetry 
as that of the MQ model, while  $H_{int}$ is invaraint under the transform $L\leftrightarrow R$ and $\alpha \leftrightarrow -\alpha$.  In the vanishing $\alpha$ limit, the time-translation invariance $G_{AB}(t_1,t_2)=G_{AB}(t_1-t_2)$ and the mirror symmetry between L and R systems $G_{RR}=G_{LL}$ and $G_{LR}(t)=-G_{RL}(t)$ can be recovered.

\section{Particle Transmission of Asymmetrically traversable wormhole}

\begin{figure}[t]
    \begin{subfigure}{0.48\linewidth}
        \centering
        \includegraphics[width=3.54in]{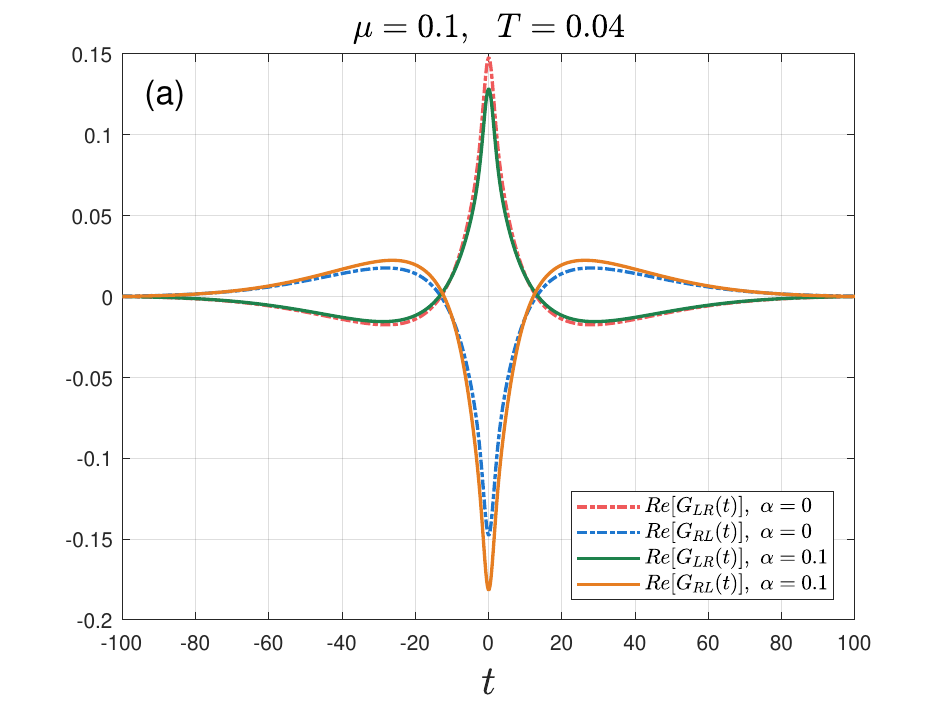}
        \label{Fig.1a}
    \end{subfigure}
    \begin{subfigure}{0.48\linewidth}
        \centering
        \includegraphics[width=3.54in]{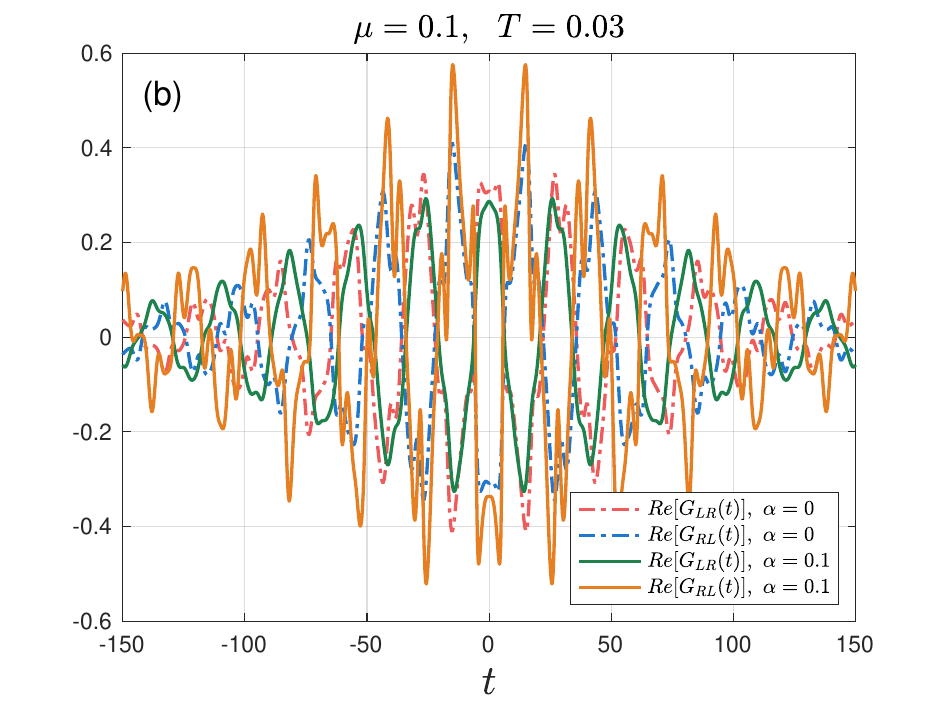}
        \label{Fig.1b}
    \end{subfigure}
    
    \caption{
    The real part of off-diagonal terms of Green's functions ${\rm Re}\left[G_{LR}(t)\right]$ and ${\rm Re}\left[G_{RL}(t)\right]$ as functions of time.  (a) In the upper half plane, the peak of ${\rm Re}\left[G_{LR}(t)\right]$ decreases as $\alpha$ increases from $\alpha=0$ to $\alpha=0.1$ in the black hole phase at temperature $T=0.04$. (b) In  the wormhole phase, the amplitude of  ${\rm Re}\left[G_{RL}(t)\right]$ amplifies at $\alpha=0.1$  at a lower temperature $T=0.03$.
    }
    \label{real_part_GF}
\end{figure}

\begin{figure}[h]
    \begin{subfigure}{0.48\linewidth}
        \centering
        \includegraphics[width=3.54in]{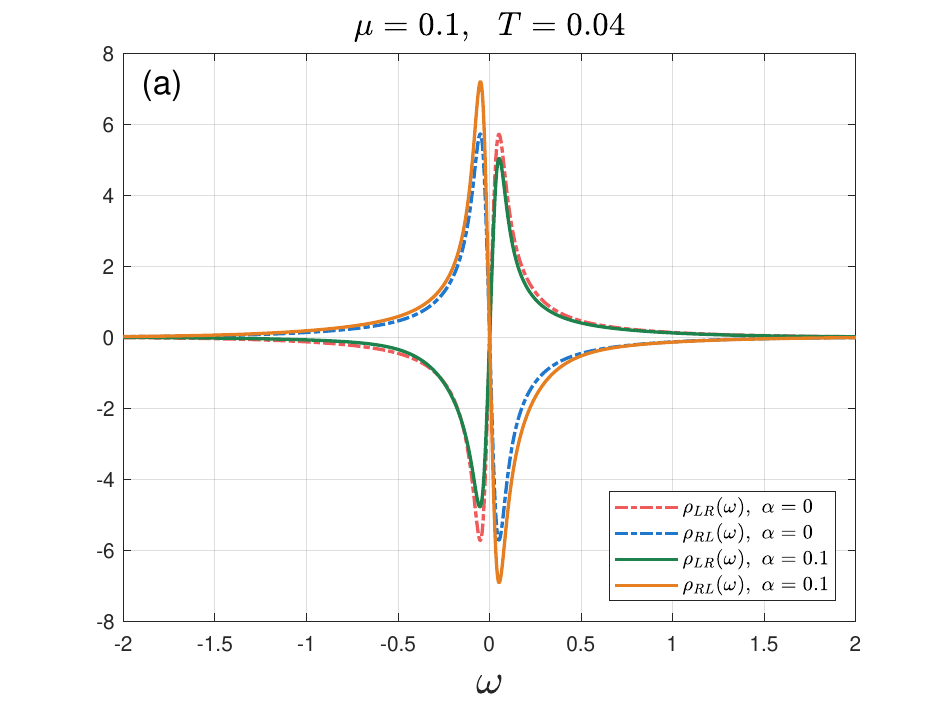}
        \label{Fig.2a}
    \end{subfigure}
    \begin{subfigure}{0.48\linewidth}
        \centering
        \includegraphics[width=3.54in]{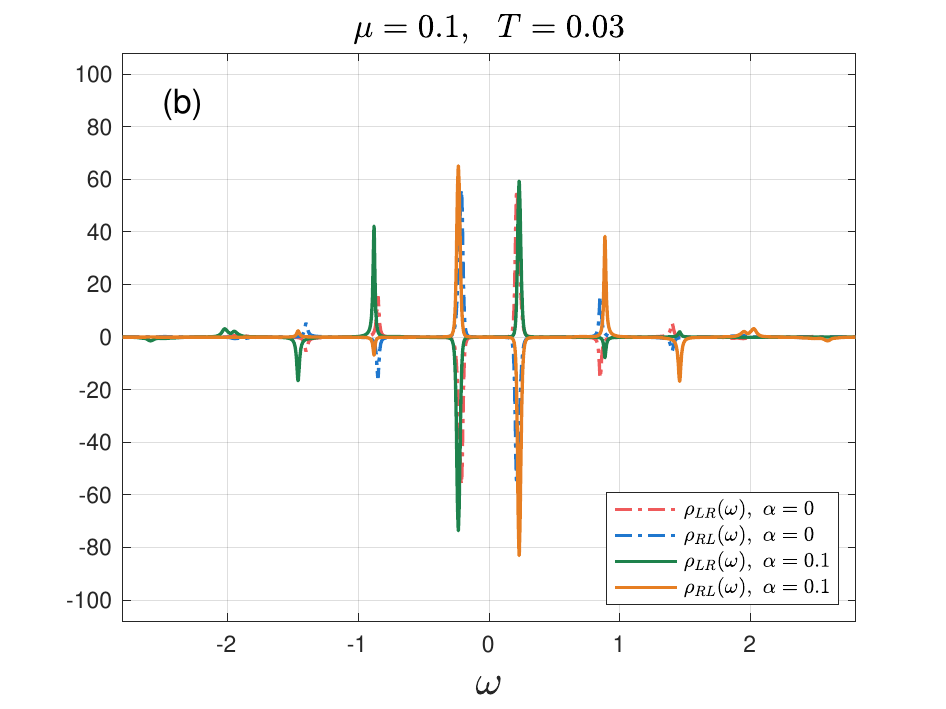}
        \label{Fig.2b}
    \end{subfigure}
    
    \caption{Spectral Functions $\rho_{LR}(\omega)$ and $\rho_{RL}(\omega)$. (a) Black hole phase at $T=0.04$. The spectral are continuous, the amplitude is obviously influenced by $\alpha$. (b) Wormhole phase at $T=0.03$. The gapped spectral functions also amplified or reduced due to the existence of $\alpha$. A striking feature is that the positions of these peaks are generally shifted by the influence of $\alpha$. As $\mu$ and $T$ decrease simultaneously, the spectral function exhibits an increment in the number of its discernible peaks.}
    \label{Spectral}
\end{figure}

\begin{figure}[h]
    \begin{subfigure}{0.48\linewidth}
        \centering
        \includegraphics[width=3.54in]{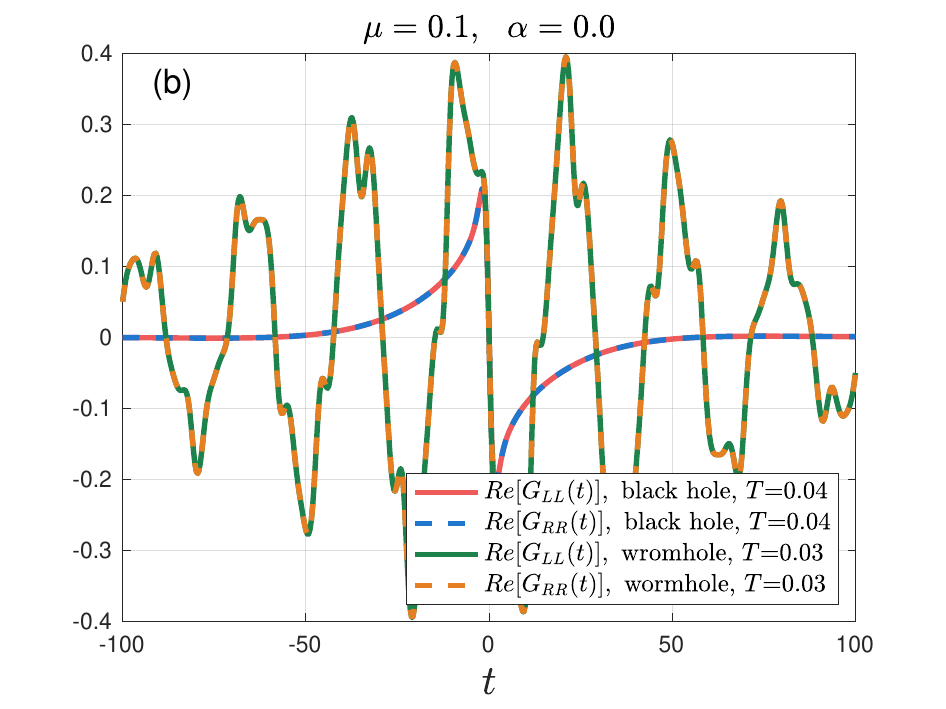}
        \label{Fig.3a}
    \end{subfigure}
    \begin{subfigure}{0.48\linewidth}
        \centering
        \includegraphics[width=3.54in]{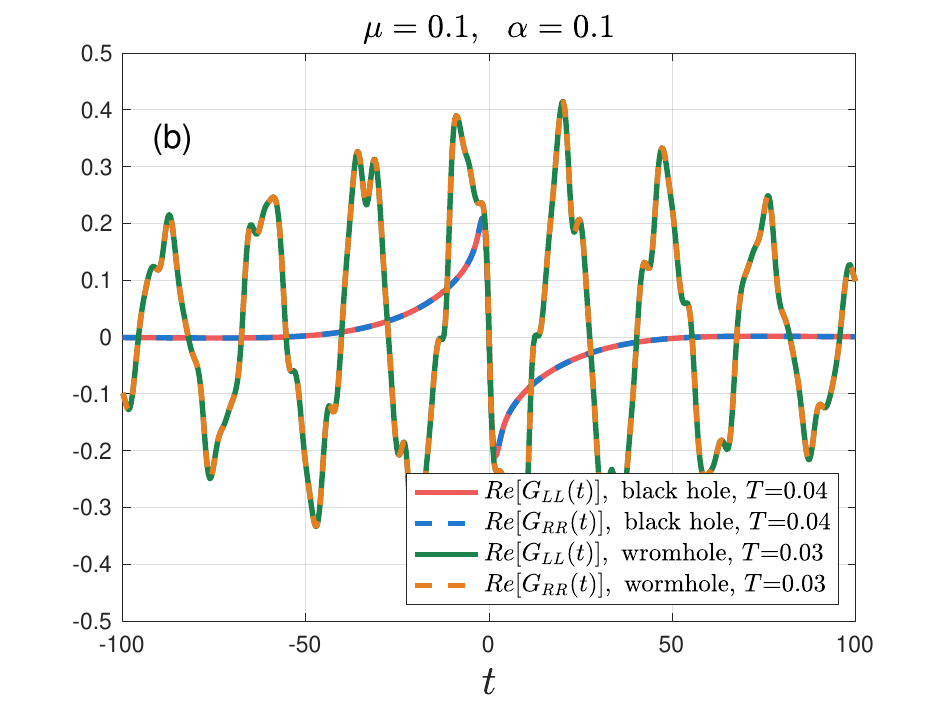}
        \label{Fig.3b}
    \end{subfigure}
        
    \caption{Diagonal terms of Green's functions $G_{LL}(t), G_{RR}(t)$ of (a) Hermitian and (b) non-Hermitian cases.}
    \label{Diagnoal_Correlators}
\end{figure}

\begin{figure}[h]
    \begin{subfigure}{0.48\linewidth}
        \centering
        \includegraphics[width=3.54in]{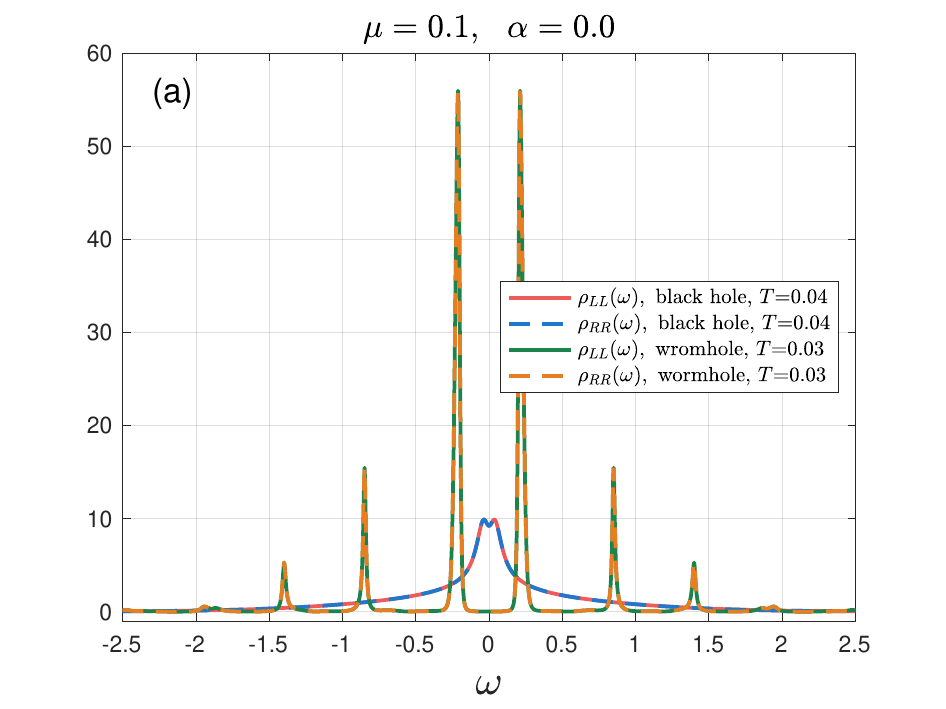}
        \label{Fig.4a}
    \end{subfigure}
    \begin{subfigure}{0.48\linewidth}
        \centering
        \includegraphics[width=3.54in]{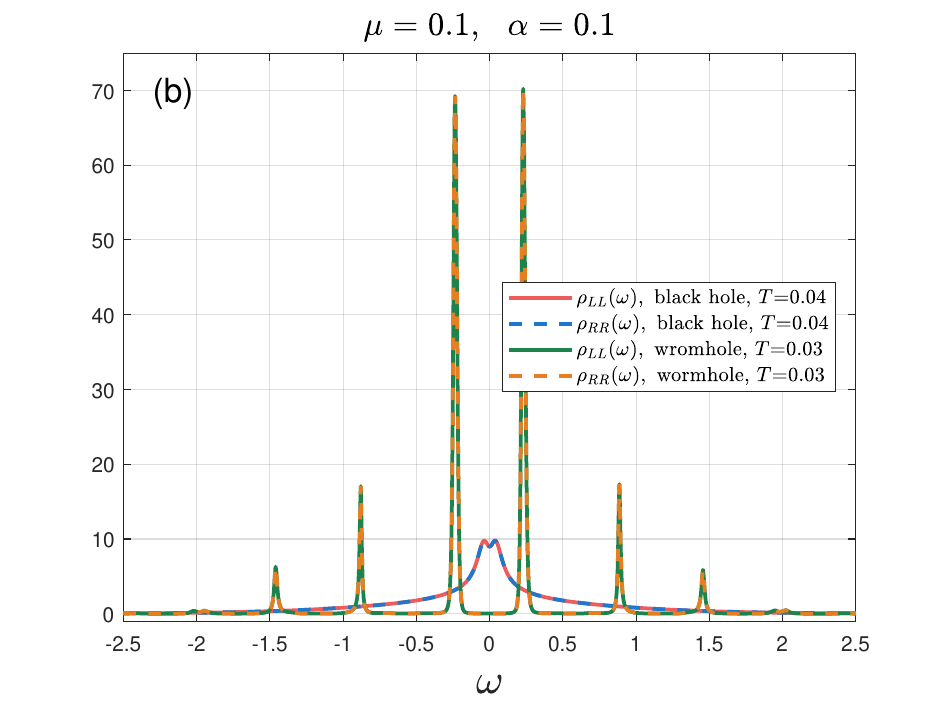}
        \label{Fig.4b}
    \end{subfigure}
        
    \caption{Spectral functions $\rho_{RR}$ and $\rho_{LL}$ of (a) Hermitian and (b) non-Hermitian cases. When $\alpha$ is non-zero, the peaks are not symmetric with respect to $\omega=0$ and all peaks become narrower and higher.}
    \label{Diagnoal_spectrum_functions}
\end{figure}
\begin{figure}[h]
    \begin{subfigure}{0.45\linewidth}
        \centering        \includegraphics[width=3.2in]{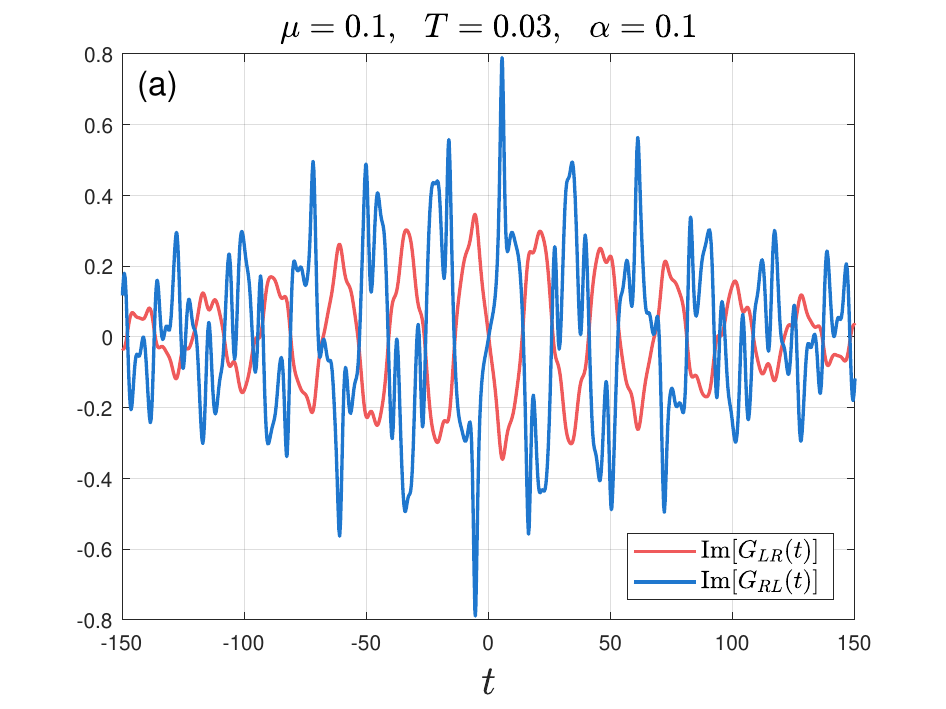}
    \end{subfigure}
    \begin{subfigure}{0.45\linewidth}
        \centering
        \includegraphics[width=3.2in]{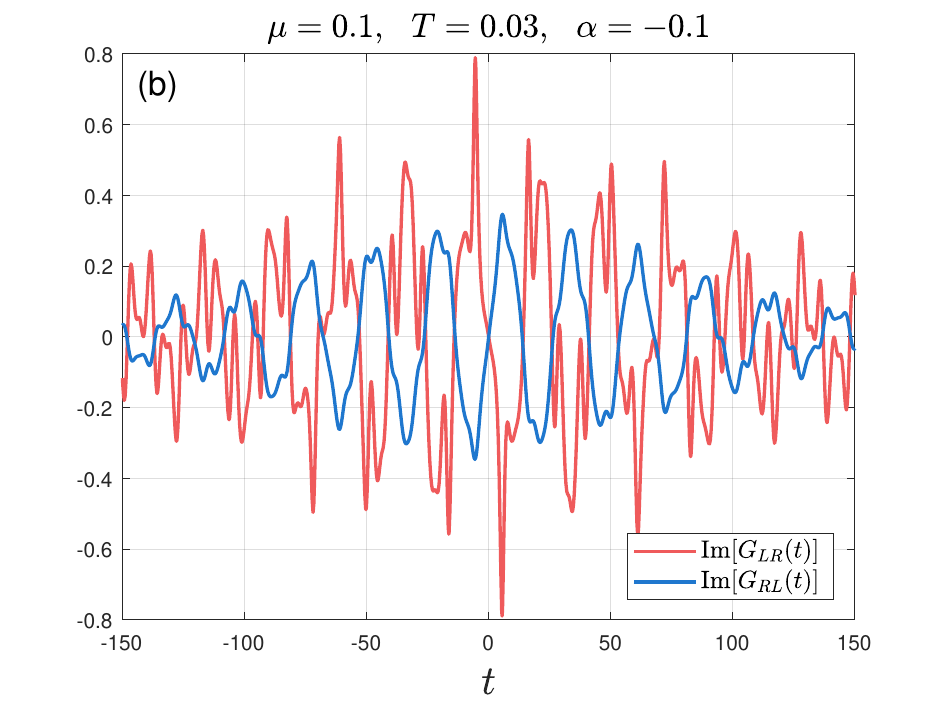}
    \end{subfigure}
    \caption{(a) The imaginary part of the off-diagonal Green functions with positive $\alpha=0.1$, the excitation created from the right side transmits more active than the excitation created from the left side;
    (b)) The imaginary part With a negative $\alpha=-0.1$, the propagation of excitation then becomes more active from left side to right side.}
    \label{fig:non-Hermitian}
\end{figure}

\begin{figure}[h]
\includegraphics[width=4.54in]{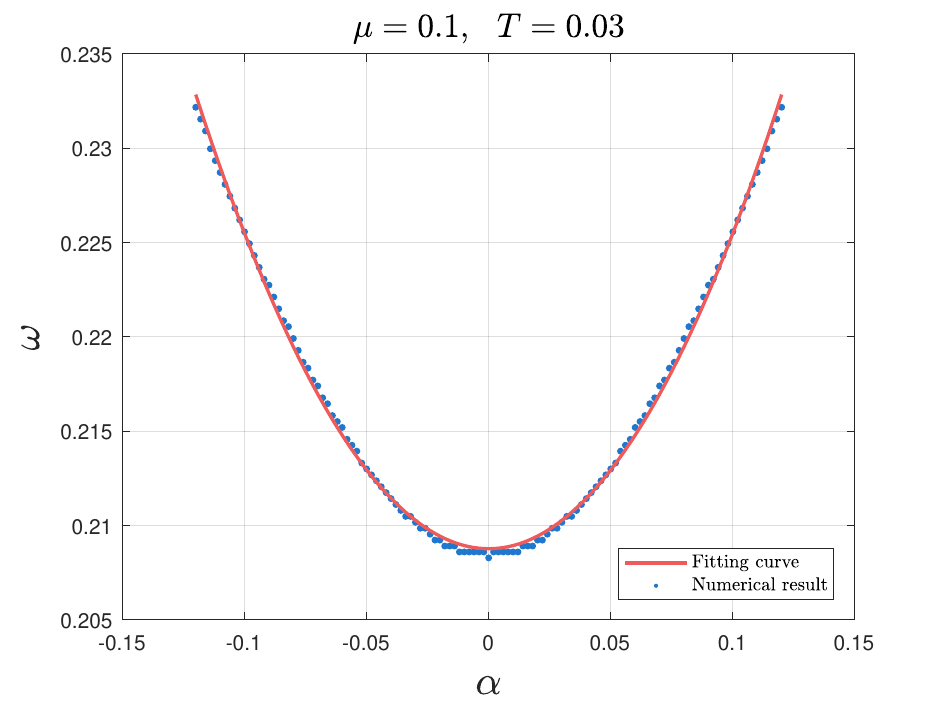}
        
    \caption{The frequency corresponding to the first peak of the spectral function $\rho_{LL}(\omega)$ (blue dots) and its quadratic fitting curve (red curve) $\omega_{\rm peak}=1.42\alpha^2+0.208$.}
\label{rhoLL_peak_w}
\end{figure}

\begin{figure}[h]
        \centering        \includegraphics[width=4.54in]{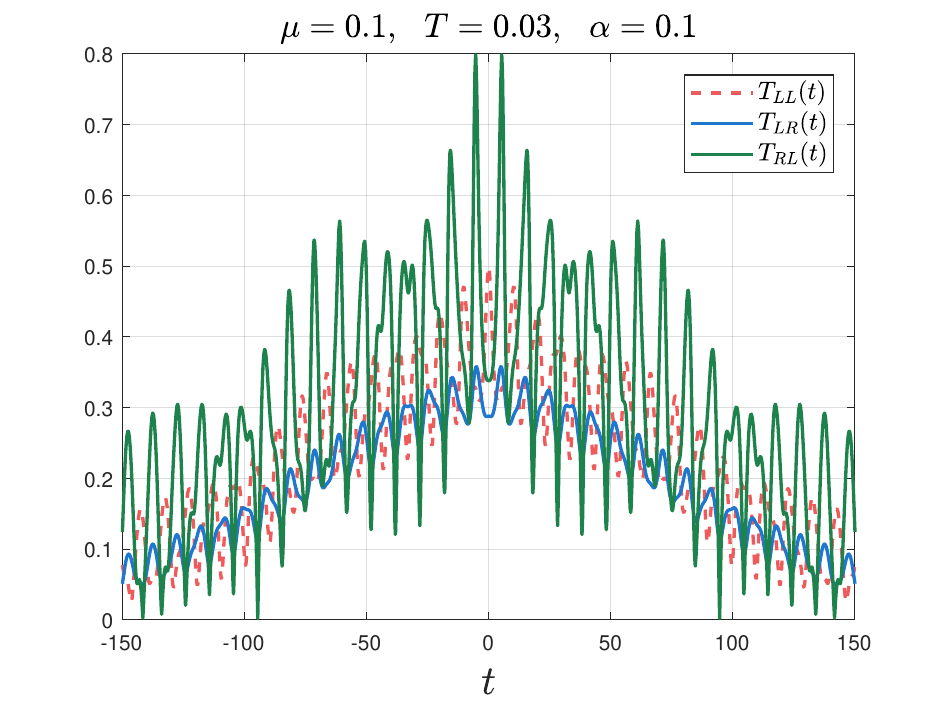}
    \caption{The non-Hermitian Transmission $T_{LL}(t)$, $T_{LR}(t)$ and $T_{RL}(t)$.} 
    \label{fig:transmission}
\end{figure}

Having derived the real-time SD equations, we strive to deeply explore the intricate dynamical properties of the non-Hermitian model by an iterative algorithm \cite{1604-07818,2003-03916,2003-03914,2009-10759}. The numerical obstacles originate from the inclusion of an exponential non-Hermitian coefficient, which complicates the SD equations significantly. Unlike the decoupling techniques utilized in \cite{2003-03914}, this complexity bars a direct simplification, thus demanding more computationally intensive calculations for non-Hermitian instances. 
To guarantee the elimination of spectral leakage or sidelobe effects, within our numerical computation framework, we have set a considerably high resolution with a discrete points count of $4 \times 10^5$ , alongside a maximum time cutoff reaching $2\times 10^3$. Our meticulous numerical results attest that both the Green's functions and spectral functions in the black hole and wormhole phases exhibit satisfactory smoothness. Moreover, the Green's functions convincingly vanish due to thermalization processes nearing the time truncation boundary.

At temperatures below the critical value ($T_{c}=0.25,~\mu_c=0.7$), as documented in \cite{2208-10800}, a clear-cut first-order phase transition occurs from a black hole phase to a wormhole phase. The overt dependency of the non-Hermitian parameter $\alpha$ on the off-diagonal real-time Green's functions  $G_{LR}$   and $G_{RL}$   is vividly displayed in Fig.\ref{real_part_GF}. Here, the distinct behaviors of these functions in each phase are separately showcased.

In the black hole phase, as seen in the upper half-plane of Fig.\ref{real_part_GF}, the real component of the left-right Green's function $G_{LR}$, exhibits a diminishing trend as the non-Hermitian parameter  $\alpha$  escalates from $\alpha=0$  to $\alpha=0.1$. Conversely, in the lower half-plane, the real part of the right-left counterpart $G_{RL}$  , rises with increasing  $\alpha$ .

When the system transits into the wormhole phase, the interaction term $H_{int}$ dominates, allowing excitations to transmit from one side to the other more active. The real parts of the off-diagonal Green's functions,  $G_{LR}$   and  $G_{RL}$, manifest intriguingly discrete attributes as depicted on the right side of Fig.\ref{real_part_GF}. These discontinuous features symbolize the capability for quantum excitation to traverse from the left-hand side to the right-hand side, indicating a non-trivial excitation transfer mechanism within the wormhole configuration. 

Moreover, in the black hole regime, the system displays a continuum spectrum, characterized by a gapless distribution of energy levels. However, upon transitioning to the wormhole phase, the spectrum experiences a qualitative shift, manifesting as the emergence of distinct, isolated energy gaps, thus creating a discontinuous energy distribution. The exact positioning of these energy gaps holds paramount importance, as it directly governs the persistence of quasiparticle excitations, where the characteristic decay rate  $\gamma$ adheres to the exponential relationship $\gamma=e^{-\beta E_{gap}/2}$.

Turning to the realm of Green's functions, it is a well-established fact that under finite-temperature conditions, the temporal evolution of Green's functions eventually decays to zero at late times. This thermodynamic decay behavior is indeed confirmed by our numerical computations. Both panels of Fig.\ref{real_part_GF}—specifically, Fig.\ref{real_part_GF}a and Fig.\ref{real_part_GF}b—attest to this attenuation pattern, visually illustrating the gradual decline in the amplitude of the Green's functions as time progresses. Based on the numerical results shown in Fig.\ref{real_part_GF}, it appears that the lifetime of the revival excitations created in a wormhole phase can significantly exceed its  duration of a black hole phase. However, it is important to note that this prolonged behavior cannot persist, especially at finite temperatures. Thermalization processes will ultimately erase all initial information, leading to the Green's function approaching zero at a sufficiently large time, denoted as $G_{AB}(t)\sim 0 $. In fact, both larger \(\mu\) and lower temperatures slow down the decay rate of amplitude of Green's functions. To quantitatively study the lifetime of the Green's function in both the wormhole phase and the black hole phase, we evaluate the time it takes for the amplitude of the Green's function to decay below a threshold of $10^{-2}$. For two phases in Fig.\ref{real_part_GF}, our numerical results indicate that in the wormhole phase, the Green's function amplitude decays to $10^{-2}$ at approximately $t\sim10^2$, while in the black hole phase , the Green's function amplitude decays to $10^{-2}$ at approximately $t\sim10$. The increased lifetime in the wormhole phase can be attributed to the discrete distribution of the spectral function. In the black hole phase, the spectral function is continuous, leading to rapid phase decoherence and faster decay of the Green's function. Conversely, in the wormhole phase, the spectral function consists of discrete peaks, resulting in sustained oscillations and a slower decay rate. This conclusion aligns with our numerical approach, where we limit our analysis to a finite truncation of the Green's function, considering times within the range of $t\in (-t_{\rm max},t_{\rm max})$.  It is worth mentioning that the interaction $H_{int}$ remains unchanged under the transformation that swaps $L$ and $R$ and simultaneously reverses the sign of $\alpha$. Derived from the definition of the Green's functions, this observation highlights the role of non-Hermitian parameters and asymmetric interactions in determining the directionality of excitation propagation. 

The portrayal of spectral functions  $\rho_{LR}(\omega)$ and $\rho_{RL}(\omega)$ in Fig.\ref{Spectral} for the black hole and wormhole phases distinctly indicates that whenever $\alpha>0$, the off-diagonal spectral functions adhere to the relation 
 $\rho_{LR}(\omega)=-e^{-4\alpha}\rho_{RL}(\omega)$. This observation signifies that the introduction of the non-Hermitian parameter $\alpha$  inherently leads to a dampening effect on the $\rho_{LR}$ component, while correspondingly amplifying the $\rho_{RL}$  component. 
 

 Fig.\ref{Diagnoal_Correlators} is the diagonal Green function of non-Hermitian wormhole and black hole terms, $G_{LL}$ and $G_{RR}$, placed in the same figure to illustrate that the diagonal terms satisfy $G_{LL}=G_{RR}$. In addition, the diagonal Green functions of the wormhole and black hole overlap at early time near $t=0$. The comparison of Hermitian and non-Hermitian cases in Fig.\ref{Diagnoal_Correlators}(a) and Fig.\ref{Diagnoal_Correlators}(b)  is used to illustrate that $ \alpha$ has minor effect on the diagonal terms.

 Fig.\ref{Diagnoal_spectrum_functions} describes the spectral function corresponding to the identical parameters as in Fig.\ref{Diagnoal_Correlators}, characterized by the positive symmetric spectral function of the diagonal Green's function. Additionally, due to the normalization condition of the spectral function, the discrete narrow spectrum corresponding to the wormhole exhibits a higher amplitude. Under the conditions of $\mu=0.1$ and $T=0.03$, there are three distinct narrow peaks, occupying 0.96 of the total area, with the first peak always dominating. From the spectral function of diagonal terms Fig.\ref{Diagnoal_spectrum_functions}, we notice that the peaks of spectral function can be affected by the non-Hermitian strength. With the existence of $\alpha$, the width of the peaks become narrower and higher.
 
Fig.\ref{fig:non-Hermitian} presents the imaginary part of the non-diagonal Green's function, which also illustrates that the enhancement and suppression of the imaginary component of the Green's function reverses when $\alpha$ takes values of $\pm0.1$. Specifically, for a positive value of $\alpha$, the magnitude of ${\rm Im} [G_{RL}(\omega)]$ is larger, whereas for a negative value, the magnitude of ${\rm Im} [G_{LR}(\omega)]$ is larger. This observation is consistent with eq.\eqref{new_mirror_symmetry}.  Meanwhile, the frequency of corresponding the maximum of peaks are also affected by $\alpha$. We further study the relationship between $\alpha$ and the position of the first peak. Fig.\ref{rhoLL_peak_w} shows the freqeuncy of first peak and its fitting curve, the numerical result indicates that its maximum frequency may square with $\alpha$ with the relation $\omega_{peak}=1.42 \alpha^2+0.208$.

The total amplitude of the Green's functions describe probability of excitation transports from sites to sites. 
To investigate the transmission probability, we can define the average site transmission amplitude through the averaged Green's functions
\begin{align}
    T_{AB}(t)=|G_{AB}(t)|.
\end{align}
Fig.\ref{fig:transmission} displays the transmission amplitudes three types of transmission amplitude $T_{LL}(t)$, $T_{LR}(t)$, $T_{RL}(t)$ in the wormhole phase. 
 For the transmission amplitudes between the left and right side $T_{LR}(t), T_{RL}(t)$, it can be observed that the excitation is first created at $t=0$ then reaches the first peak after half a period $t=\frac{1}{2}t_{re}$ and hits the bottom  $t=t_{re}\sim \mu^{2/3}$ after another half period. 
 The recurrence time $t_{re}$ indicates stronger coupling leads to denser oscillations. The Green's function oscillates with a period proportional to $\mu^{-2/3}$, explained by the energy gap in Euclidean time. This gap leads to exponential decay of Green's functions $G_{AB}(t) \propto e^{-E_{gap}\tau}$, resulting in an oscillating frequency $\omega_{re} = \mu^2/3$. Previous work \cite{2208-10800} studied decay rates and numerically demonstrated the energy gap's proportionality to $\mu^{-2/3}$.
In our study, we have thoroughly investigated the dynamics of non-Hermitian SYK models. For a comprehensive understanding of the behavior of Green's functions and spectral functions under various parameter settings, we have included additional numerical analysis in Appendix \ref{appB}. This supplementary analysis provides further insights into the system's response under different temperature and non-Hermitian parameter configurations.

\section{Out-of-time-ordered correlators}\label{Finite_N}
In the subsequent analysis, we embark on an exhaustive numerical exploration of this non-Hermitian two coupled SYK model, confined to a finite system size, with the express purpose of unraveling its intricate chaotic behavior and the accompanying emergent features. The Jordan-Wigner transformation allow us to build the exact matrix of the Hamiltonian through a spin-chain, the specific form we set is
\begin{align}
    C_i=\sigma_i^-\bigotimes_{j>i}\sigma^z_j~,\quad
    C_i^\dagger=\sigma_i^+\bigotimes_{j>i}\sigma^z_j~,
\end{align}
where the $\sigma_i^{\pm}, \sigma_i^{z}$ represent the spin Pauli matrices at each site, different sites operator are connected by the tensor product. Every single complex fermion lives in a $2\times 2$ dimension Fock space. For the case of two-coupled SYK model, the number of the fermion of both side are $N$, therefore the Hamiltonian takes $4^N\times 4^N$ dimensions.  The diagonalization of the matrix representation of Hamiltonian \eqref{hamilton} permits a direct extraction of the energy spectrum. Nevertheless, the dimension of this matrix grows exponentially with the number of lattice sites, presenting formidable computational hurdles for exact diagonalization methods. In light of the stringent constraints imposed by current computational resources, the practical scope for undertaking such a task is necessarily restricted to a meticulously chosen, finite range of system sizes, parameterized by $N$. For the present study, we have opted to carry out numerical computations for the exemplary case of $N=6$, cognizant of the fact that the computational and memory requirements swiftly outstrip the capacities of conventional computational platforms. 

\begin{figure}[h]
        \centering
        \includegraphics[width=4.54in]{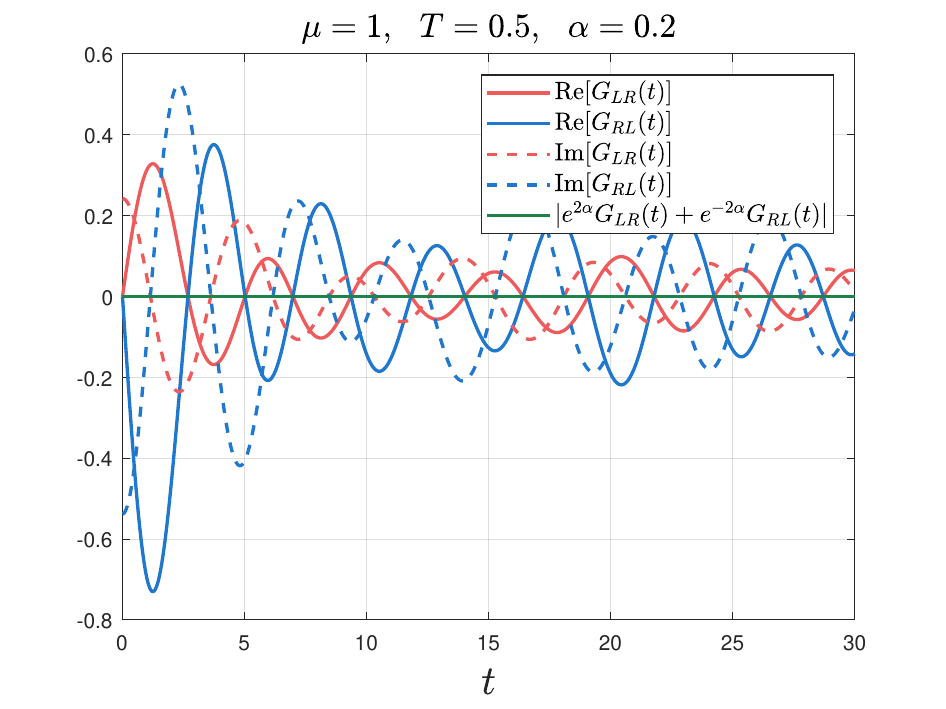}

    \caption{The $N=6$ full diagonalization implement of Green's function. With positive $\alpha$, $G_{RL}(t)$ is amplified and $G_{LR}(t)$ is still suppressed, which is consistent with the result of large $N$ calculation. Their difference $|e^{2\alpha}G_{LR}(t)+e^{-2\alpha}G_{RL}(t)|$ at $N=6$ is shown as the green solid line which is vanished.}
    \label{ED_Correlator}
\end{figure}

The exact diagonal  method offers an alternative approach to re-examine the relation of off-diagonal correlators relation \eqref{new_mirror_symmetry}. The averaged two point function is defined as 
\begin{equation}
G_{AB}(t)=\sum_{i=1}^N\frac{{\rm Tr}\left[e^{-\beta H} C_{i}^{A}(t)C_{i}^{B}(0)\right]}{N~{\rm Tr}{\left[e^{-\beta H}\right]}}.
\end{equation}
It is crucial to emphasize that when dealing with a non-Hermitian Hamiltonian, the inverse time-evolution operator fundamentally deviates from being unitary. Mathematically, this can be expressed as  $U(-t)=e^{iHt}\neq U(t)^\dagger=e^{iH^\dagger t}$, due to the non-commutation relation  $e^{iH^\dagger t}e^{-iHt}\neq 1$. This inherent departure from unitarity has profound implications for the ED method, necessitating the independent computation of the exponentiated matrix corresponding to the inverse time evolution. Figure \ref{ED_Correlator} serves as a testament to the robustness of \eqref{new_mirror_symmetry} even for finite-$N$ correlators. The numerical data, depicted by the green solid line, attests to an exceptionally tight agreement, with the maximum absolute deviation satisfying  ${\rm Max}|e^{2\alpha}G_{LR}(t)+e^{-2\alpha}G_{RL}(t)|<10^{-15}$. This near-perfect alignment can be confidently attributed to numerical precision limits rather than genuine deviations from the predicted symmetry, thereby providing strong empirical validation for the theoretical framework under scrutiny.

We are now able to numerically implement a finite-size non-Hermitian SYK model, which allows us to initiate an initial exploration of the system's chaotic behavior. In a quantum system, the chaotic behaviour can be described by the exponential growth behaviour of the OTOCs. In order to mitigate the impact of finite-size effect, we choose the specific form of OTOCs to be the 'regularized' OTOCs.

\begin{align}
F(t)=\frac{\overline{\trace{\left[ W_i(t)\rho^{1/4}V_j(0)\rho^{1/4}W_i(t)\rho^{1/4}V_j(0)\rho^{1/4}\right]}}}{\trace{ \rho}},
\end{align}
where $\rho=e^{-\beta H}$ and the operators $W_i$ and $V_j$ are chosen to be the fermion operators from different sites $W_i=C_i^A+C_i^{A\dagger}$,$V_j=C_j^B+C_j^{B\dagger},~ i\neq j$, the index $A,B=L,R$ are represented for the left or right side fermion operators. The over line represents the average of the random ensembles. This specific form has been detailed investigated and demonstrated in \cite{2002-05725,PhysRevResearch.2.013254} that it can capture the behaviour of the large $N$ limit condition better and more reliably compared to the unregularized form 
\begin{equation}
    F^{(u)}(t)=\overline{\braket{ W(t)V(0)W(t)V(0)}}_{\beta}.
\end{equation}

\begin{figure}[h]
    
        \centering
        \includegraphics[width=4.54in]{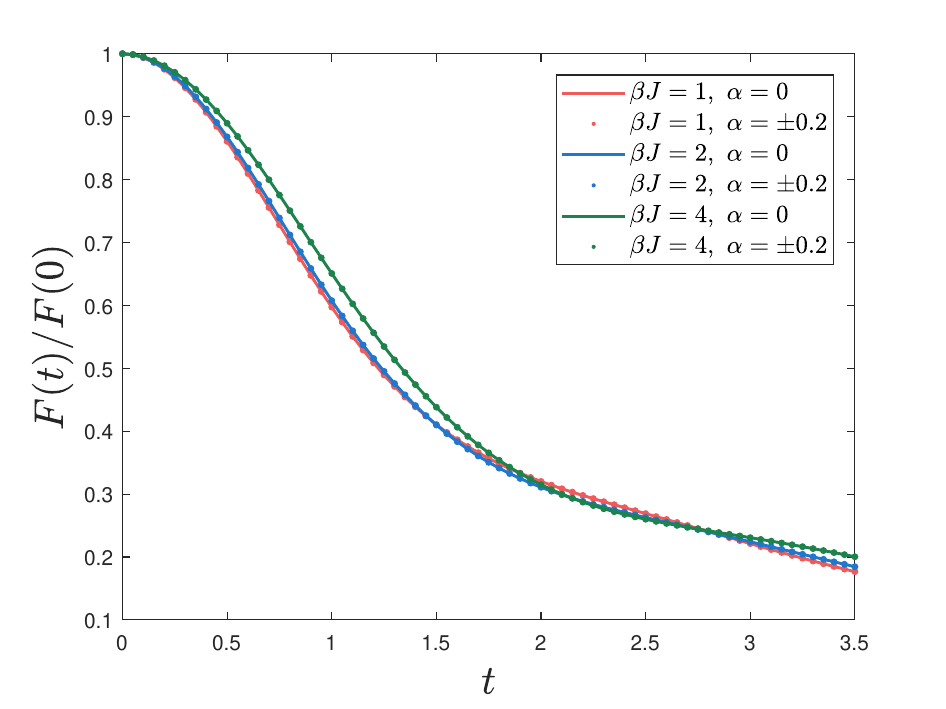}

    \caption{The averaged regularized OTOCs with 50 implements average, where we chose $N=6$ and $\mu=0.05$.  The graphical representation illustrates that regardless of the choice of $\alpha$, the regularized OTOCs display insensitivity to it, as evidenced by the parallel trends in the non-Hermitian (depicted by dotted lines) and Hermitian (represented by solid lines) scenarios. Upon lowering the temperature, a notably slower decline in slope during the initial time period emerges, suggesting a diminishing Lyapunov exponent with decreasing temperature – a trend that becomes more pronounced as temperatures drop.}
    \label{OTOC}
\end{figure}

To study how exactly does the non-Hermitian model differ from the Hermitian case in the behaviour of OTOCs, we calculate the OTOCs of different $\alpha$ cases, as shown in fig.\ref{OTOC}. Much of the earlier literature \cite{garcíagarcía2024lyapunov,PhysRevResearch.2.013254, 2002-05725, PhysRevB.96.054503} suggests the regularized OTOCs obey a leading exponential growth behaviour by $e^{\lambda t}/N$. It is is well represented in the early time behavior of OTOCs in Fig.\ref{OTOC}. This implies that one can extract the Lyapunov exponents by capturing the early segments in OTOCs that exhibit exponential growth. As shown in Fig.\ref{OTOC}, we calculate the fifty implements average for the regularized OTOCs, setting the operators $W_{i}$ and $V_j$ on the same side to explore the behavior of the chaos exponent. Our numerical result shows that once the operator are fully single-sided, the $F(t)$ is independent of $\alpha$ no matter positive or negative. Furthermore, $F(t)$ is same for both case: $W_i=C_i^L+C_i^L, V_i=C_j^L+C_j^L$ and $W_i=C_i^R+C_i^R, V_i=C_j^R+C_j^R$. This is partially due to the fact that the chaos behaviour of SYK model is driven by the random interaction whereas the four body interactions on both the left and right sides share the exact same Gaussian random variables. At early time, $F(t)$ shows a exponential decay as expect and the decay rate is inversely proportional to $\beta J$ which implies a smaller Lyapunov exponent as temperature goes down.

In \cite{2002-05725}, the authors conducted a more detailed and extensive study on many-body chaos in the single-sided Majorana SYK model. They used Krylov subspace method to investigate regularized OTOCs in extremely large and sparse matrices. This method leverages the sparsity of the Hamiltonian to avoid excessive memory usage and significantly accelerates the time evolution of states through the Krylov subspace algorithm. With these optimizations, it becomes feasible to calculate the Lyapunov exponents for systems with much larger number of fermions. They computed the Lyapunov exponents for up to 60 Majorana fermions, the results shows that the Lyapunov exponents matched well with the large \( N \) results obtained from SD equations as expectation. This indicates that both for the finite size SYK model with \( N = 60 \) and JT gravity satisfy the chaos bound, which is a necessary requirement for gravitational duals  as explained by \cite{1604-07818}. This finding is also highly relevant to the many-body chaos behavior exhibited by non-Hermitian OTOCs. While such algorithmic optimization requires comprehensive consideration and thorough investigation for non-Hermitian cases, our current approach still uses matrix exponential of the time evolution operator to compute OTOCs. Although this method is not as efficient as the Krylov subspace algorithm, it guarantees more accurate and reliable results. It also suggests that the feasibility of Krylov subspace algorithm in non-Hermitic cases is a topic worth studying.

Specifically, as shown in Fig.\ref{OTOC}, the initial exponential growth phase of the OTOCs demonstrates a slower rate with decreasing temperatures, indicating smaller Lyapunov exponents. This behavior is consistent with the findings in the literature, suggesting that even with the computational constraints, our results are in good agreement with the established behavior. 

These observations are crucial for understanding the implications of maximal chaos and its gravitational duals in non-Hermitian systems. Therefore, despite the limitations imposed by the small system size, our findings align well with larger system studies and provide a robust foundation for further exploration of chaotic dynamics in non-Hermitian SYK models.

\section{Conclusion and discussions}
In summary, we investigate the revival dynamics of excitations in a coupled non-Hermitian SYK system, uncovering a novel connection between diagonal and off-diagonal Green's functions (Eq. (11)). Despite directional asymmetry in propagation (right-to-left vs. left-to-right), the system retains distinct black hole and wormhole phases. Notably, the black hole phase features a gapless continuous energy spectrum, whereas the wormhole phase introduces isolated energy gaps critical for quasiparticle longevity, with decay rate 
$\gamma \propto e^{-\beta E_{gap}/2}$. Finite-temperature Green's function simulations robustly demonstrate thermally-driven late-time decay, as visually exemplified in Fig.\ref{real_part_GF}a and b.

This study also examines the spectral functions 
$\rho_{RR}$ and $\rho_{LL}$  associated with Hermitian and non-Hermitian systems, respectively, revealing that when the parameter $\alpha$
 assumes a non-zero value, the symmetry of the peaks with respect to $\omega=0$ is broken, and all peaks exhibit increased height and reduced width. The spectral functions $\rho_{LR}$ and $\rho_{RL}$are analyzed for both the black hole phase at $T=0.04$ and the wormhole phase at $T=0.03$. In the black hole phase, the spectra remain continuous, with the amplitude significantly influenced by $\alpha$. In the wormhole phase, the gapped spectra also experience amplification or reduction due to the presence of $\alpha$, and a notable observation is that the positions of these peaks are systematically shifted under the effect of $\alpha$. As both coupling $\mu$ and temperature $T$ decrease concomitantly, the spectral function displays an increase in the number of discernible peaks.

Investigating the off-diagonal terms of Green's functions, $G_{RL}(t)$ and $G_{LR}(t)$  for the Hermitian and non-Hermitian cases, respectively, reveals that when $\alpha$
is positive (e.g., 
$\alpha=0.1$), excitations originating from the right side transmit more actively than those from the left. Conversely, with a negative $\alpha=-0.1$, excitation propagation is more pronounced from left to right. The frequency corresponding to the first peak of the spectral function $\rho_{LL} (\omega)$
 is found to follow a quadratic dependence on $\alpha^2$, specifically $\omega_{\rm peak}=1.42\alpha^2+0.208$. The non-Hermitian transmission coefficients $T_{RR}(t)$, $T_{LR}(t)$ and $T_{RL}(t)$ are also explored. We have meticulously verified the symmetry properties fulfilled by the off-diagonal Green's functions via the finite $N$ ED methodology. Notably, even for finite $N$ scenarios, OTOCs persistently exhibit exponential decay which is independent of non-Hermiticity, a finding that holds true for both Hermitian and non-Hermitian system realms. This exhaustive scrutiny provides profound insights into the complex revival dynamics governing non-Hermitian systems, along with their subtle phase transition characteristics

\acknowledgments
 We would like to thank Wenhe Cai, Cheng Peng, Antonio M.García-García, Zhuoyu Xian and Zhenbin Yang for helpful discussions. We are especially grateful to Jinwu Ye for  valuable comments on the manuscript. This work is partly supported by NSFC, China (Grant No. 12275166 and  No. 12311540141). 

\bibliographystyle{apsrev4-1}
\bibliography{refs.bib}

\appendix

\section{Derivation of Schwinger-Dyson equations}\label{appendix1}

In this appendix, we will start from the averaged partition functions to derive the corresponding SD equations. For simplicity, we first employ the Euclidean time signature, one can obtain the real-time formalism through the wick rotation. After applying the replica-trick the M-copy replicated partition functions under averaged over the Gaussian random variables reads
\begin{align}
\overline{Z^M} &= \int\prod_{i,A} \mathcal{D}C_i^{A\dagger} \mathcal{D}C_i^A ~\prod_{i,j,k,l} dJ_{ijkl} dJ_{ijkl}^* e^{-|J_{ijkl}|^2/\sigma^2} \exp\left[-\sum_{m=1}^M\int d\tau_1 \sum_{i,A} C_{im}^{A\dagger} \partial_\tau C_{im}^A \right. \nonumber\\
& \left. +i\mu\sum_{i}(e^{-2\alpha}C_{im}^{L\dagger}C_{im}^R-e^{2\alpha}C_{im}^{R\dagger}C_{im}^L)-J_{ijkl}
\sum_{i,j,k,l,A}( 2C_{im}^{A\dagger} C_{jm}^{A\dagger} C_{km}^A C_{lm}^A + 4C_{im}^{A\dagger} C_{jm}^L C_{km}^{L\dagger} C_{lm}^A
) \right],
\end{align}
where the second lower index $m$ of fermions $C_{im }$ denotes the $m$-th replica copy. After integrating out all Gaussian random integrals, the replicated average partition becomes
\begin{align}
\overline{Z^M} &= \int\prod_{i,A} \mathcal{D}C_i^{A\dagger} \mathcal{D}C_i^A \exp\left[-\sum_{m=1}^M\int d\tau_1 \sum_{i,A} C_{im}^{A\dagger} \partial_\tau C_{im}^A +i\mu\sum_{i}(e^{-2\alpha}C_{im}^{L\dagger}C_{im}^R-e^{2\alpha}C_{im}^{R\dagger}C_{im}^L)\right. \nonumber\\
& \left.+9NJ^2\sum_{m,n}\int d\tau_1 d\tau_2 \sum_{A,B=L,R}(\frac{1}{N}\sum_{i}C_{im}^{A}(\tau_2)C_{in}^{B\dagger}(\tau_1))^2(\frac{1}{N}\sum_iC_{im}^{B}(\tau_1)C_{in}^{A\dagger}(\tau_2))^2\right].
\end{align}
Next, insert the identity
\begin{align}
    \int \mathcal{D} \Sigma\mathcal{D}G \exp \left[ N\int d\tau_1 d\tau_2\sum_{A,B}\Sigma^{AB}_{mn}(\tau_1,\tau_2)\left(G^{BA}_{mn}(\tau_1,\tau_2) - \frac{1}{N} \sum_i C_{im}^{B}(\tau_1) C_{in}^{A\dagger}(\tau_2) \right)
 \right]=1
 ,
\end{align}
this is equivalent to introduce two auxiliary functions to package the fermion pair $\frac{1}{N}\sum_iC_{im}^{A}(\tau_1)C_{in}^{B\dagger}(\tau_2)$. The partition functions can be rewritten as
\begin{align}
\overline{Z^M} &= \mathcal{D}\Sigma\mathcal{D}G  \exp\left[ - \sum_{m=1}^M\int d\tau_1 d\tau_2 \delta(\tau_2-\tau_1)\left(\sum_{i,A} C_{im}^{A\dagger}(\tau_2) \partial_{\tau_1} C_{im}^A(\tau_1) \right. \right. \nonumber\\
&\left. \left. +i\mu\sum_{i}\left(e^{-2\alpha}C_{im}^{L\dagger}(\tau_2)C_{im}^R(\tau_1)-e^{2\alpha}C_{im}^{R\dagger}(\tau_2)C_{im}^L(\tau_1)\right) \right) \right. -\int d\tau_1 d\tau_2 \sum_{A,B}\sum_{m,n}C_{im}^A(\tau_2)\Sigma_{mn}^{AB}(\tau_1,\tau_2)C_{in}^{B\dagger}(\tau_1)\nonumber\\
&\left. +NJ^2\sum_{m,n}\int d\tau_1 d\tau_2 \sum_{A,B}\Sigma_{mn}^{AB}(\tau_1,\tau_2)G_{mn}^{BA}(\tau_2,\tau_1)+9\sum_{A,B}G_{mn}^{AB}(\tau_2,\tau_1)^2 G_{mn}^{BA}(\tau_1,\tau_2)^2 \right].
\end{align}
As illustrated in \cite{softmode}, the most natural replica-diagnoal solotion is $\Sigma_{mn}^{AB}=\Sigma^{AB}\delta_{mn}$. The form with only replica diagonal terms significantly simplifies the action, now the partition function is simplified as 
\begin{align}
\overline{Z^M} &= \int\mathcal{D}\Sigma\mathcal{D}G e^{-N S^{(M)}} = \int\mathcal{D}\Sigma\mathcal{D}G \exp\left[ - \sum_{m=1}^M\int d\tau_1 d\tau_2 \delta(\tau_2-\tau_1)\left(\sum_{i,A} C_{im}^{A\dagger}(\tau_2) \partial_{\tau_1} C_{im}^A(\tau_1) \right. \right. \nonumber\\
&\left. \left. +i\mu\sum_{i}\left(e^{-2\alpha}C_{im}^{L\dagger}(\tau_2)C_{im}^R(\tau_1)-e^{2\alpha}C_{im}^{R\dagger}(\tau_2)C_{im}^L(\tau_1)\right) \right) \right. -\int d\tau_1 d\tau_2 \sum_{A,B}\sum_{m}C_{im}^A(\tau_2)\Sigma^{AB}(\tau_1,\tau_2)C_{im}^{B\dagger}(\tau_1)\nonumber\\
&\left. +NJ^2\sum_{m}\int d\tau_1 d\tau_2 \sum_{A,B}\Sigma^{AB}(\tau_1,\tau_2)G^{BA}(\tau_2,\tau_1) +9\sum_{A,B}G^{AB}(\tau_2,\tau_1)^2 G^{BA}(\tau_1,\tau_2)^2 \right].
\end{align}
Recast the remaining fermionic terms into a determinant and utilize the replica trick
\begin{align}
-\overline{\ln Z}=-\lim_{M\rightarrow 0}\frac{\ln \overline{Z^M}}{M},
\end{align}
the action $-\overline{\ln Z}$ is eventually expressed as the saddle point of the integral $\mathcal{D}\Sigma,~\mathcal{D}G$
\bea
\frac{S}{N}=-\ln \det \mathcal{M}+\int d\tau_1 d\tau_2 \sum_{A,B}\Sigma_{AB}(\tau_1,\tau_2)G_{BA}(\tau_2,\tau_1) +9J^2\sum_{A,B}G_{AB}(\tau_2,\tau_1)^2 G_{BA}(\tau_1,\tau_2)^2.
\eea
where we finally put the left and right label $A,B=L,R$ on subscript to make them consistent with the text. The matrix $\mathcal{M}$ in Matsubara frequency representation reads
\bea
\mathcal{M} = \bigoplus_{n} \begin{pmatrix}
-i\omega_n - \Sigma_{RR} & i\mu e^{2\alpha} + \Sigma_{RL} \\
-i\mu e^{-2\alpha}+ \Sigma_{LR} & -i\omega_n - \Sigma_{LL}
\end{pmatrix}.
\eea
The time translation invariance indicates $G_{AB}(\tau_1,\tau_2)=G_{AB}(\tau_1-\tau_2)=G(\tau)$. The functional variation, for example, $\frac{\delta S(\Sigma,G)}{\delta G_{LR}(\tau^\prime)}$ is
\begin{align}
0=\frac{\delta S[\Sigma,G]}{\delta G_{LR}(\tau^\prime)}&=\beta\int d\tau \Sigma_{RL}(\tau)\delta_{\beta}(\tau^\prime+\tau) +18J^2G_{LR}(-\tau) G_{RL}(\tau)^2\delta_\beta(\tau^\prime+\tau)
+18J^2G_{LR}(\tau) G_{RL}(-\tau)^2\delta_\beta(\tau^\prime-\tau)\nonumber\\
&=\Sigma_{RL}(-\tau^\prime)+36J^2G_{LR}(\tau^\prime) G_{RL}(-\tau^\prime)^2,
\end{align}
where the $\delta_\beta(\tau)$ is a counterperiodic Dirac comb with the period of $\beta$. Move the $\Sigma_{RL}$ to the left hand side and remark $\tau^\prime\rightarrow -\tau$ we derive
\bea
\Sigma_{RL}(\tau)=-36J^2 G_{RL}(\tau)^2 G_{LR}(-\tau).
\eea
After the functional variation over all auxilary fields $G_{AB}, \Sigma_{AB}$ and applying the Wick rotation, we ultimately obtain the real-time SD equations \eqref{SDequation}. Specifically, for ${\delta S(\Sigma,G)}/{\delta G_{AB}}=0$, this corresponds to
\bea
\Sigma^>_{AB}(t)=-36J^2G^{>2}_{AB}(t
)G^>_{BA}(-t),
\eea
for ${\delta S(\Sigma,G)}/{\delta \Sigma_{AB}}=0$ we obtain
\bea
&\tilde{G}^R_{LL}(\omega,\alpha)=\frac{-\omega-\tilde{\Sigma
}^R_{RR}(\omega,\alpha)}{\tilde{D}(\omega,\alpha)},\nonumber\\
&\tilde{G}^R_{RR}(\omega,\alpha)=\frac{-\omega-\tilde{\Sigma
}^R_{LL}(\omega,\alpha)}{\tilde{D}(\omega,\alpha)},\nonumber\\
&\tilde{G}^R_{LR}(\omega,\alpha)=\frac{-i\mu e^{-2\alpha}+\tilde{\Sigma
}^R_{LR}(\omega,\alpha)}{\tilde{D}(\omega,\alpha)},\nonumber\\
&\tilde{G}^R_{RL}(\omega,\alpha)=\frac{i\mu e^{2\alpha}+\tilde{\Sigma
}^R_{RL}(\omega,\alpha)}{\tilde{D}(\omega,\alpha)},\nonumber\\
&\tilde{D}(\omega,\alpha)=\left( -\omega-\tilde{\Sigma
}^R_{LL} \right)\left( -\omega-\tilde{\Sigma
}^R_{RR} \right)+\left( i\mu e^{-2\alpha}-\tilde{\Sigma
}^R_{LR} \right)\left( i\mu e^{2\alpha}-\tilde{\Sigma
}^R_{RL} \right).
\eea

\section{Additional Numerical analysis for different \texorpdfstring{$\alpha$}{alpha} and \texorpdfstring{$\mu$}{mu}}\label{appB}

\begin{figure}[h]
\begin{subfigure}{0.45\linewidth}
        \centering        \includegraphics[width=3.2in]{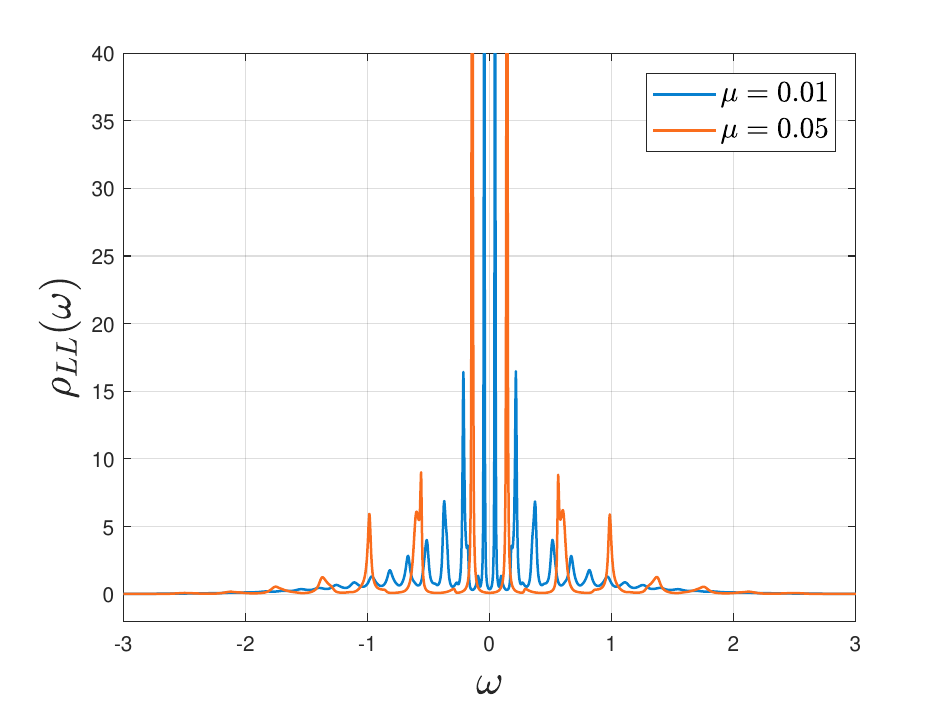}
        \caption{}
    \end{subfigure}
    \begin{subfigure}{0.45\linewidth}
        \centering
        \includegraphics[width=3.2in]{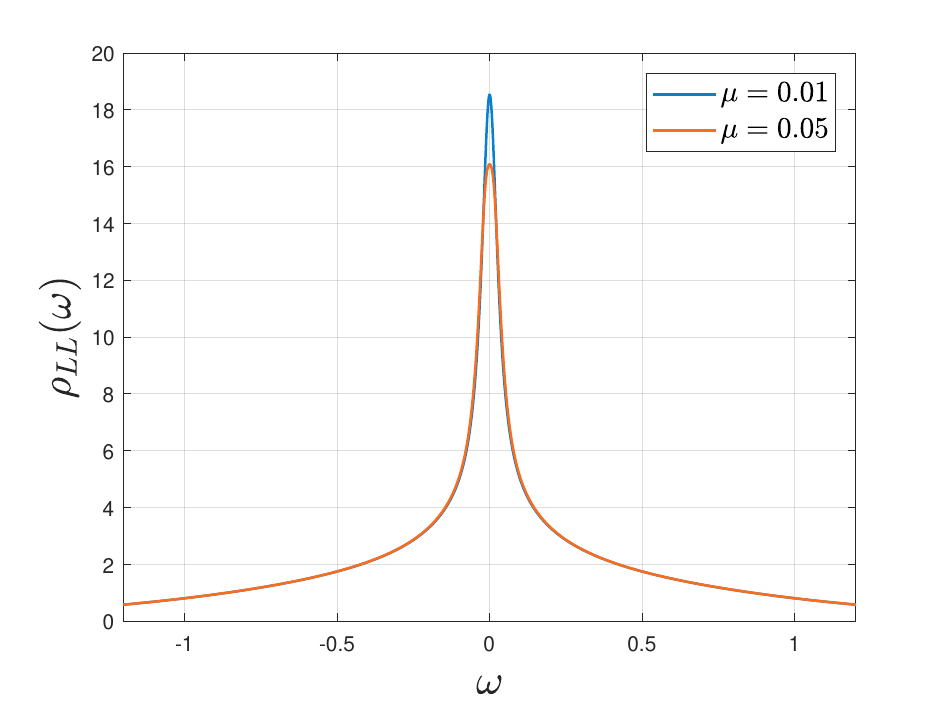}
        \caption{}
    \end{subfigure} 
    \caption{(a) The wormhole phase spectral functions at lower temperature, where $\alpha=0.05, T=0.002, \mu=0.01, 0.05$. (b)  The black hole phase spectral functions at lower temperatures, where $\alpha=0.05, T=0.02, \mu=0.01, 0.05$.}
\label{rho_small_mu}
\end{figure}
\begin{figure}[h]
\begin{subfigure}{0.45\linewidth}
        \centering        \includegraphics[width=3.2in]{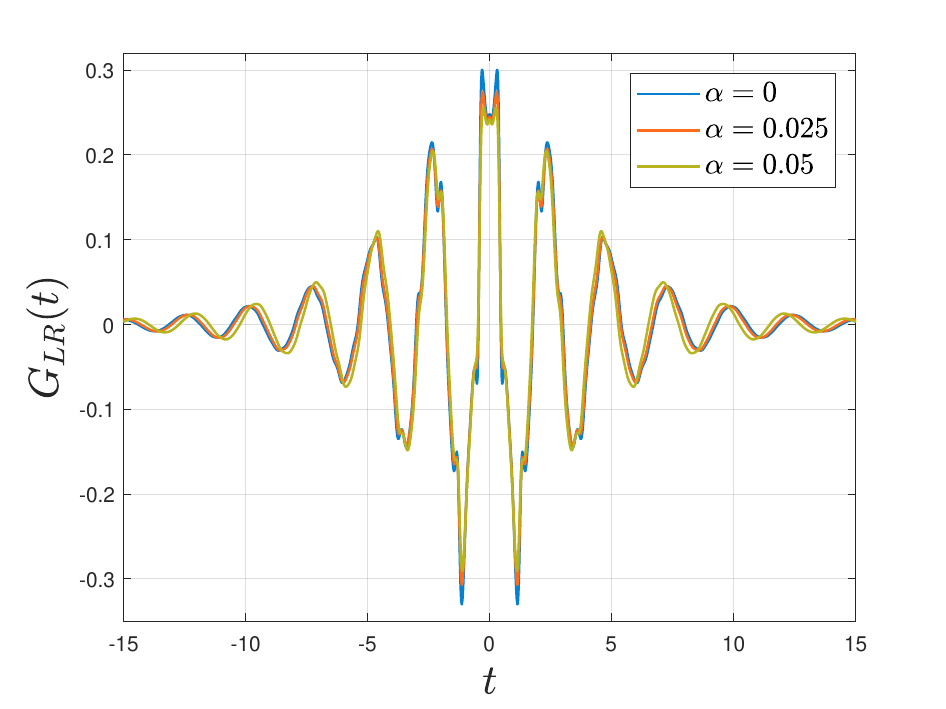}
        \caption{}
    \end{subfigure}
    \begin{subfigure}{0.45\linewidth}
        \centering
        \includegraphics[width=3.2in]{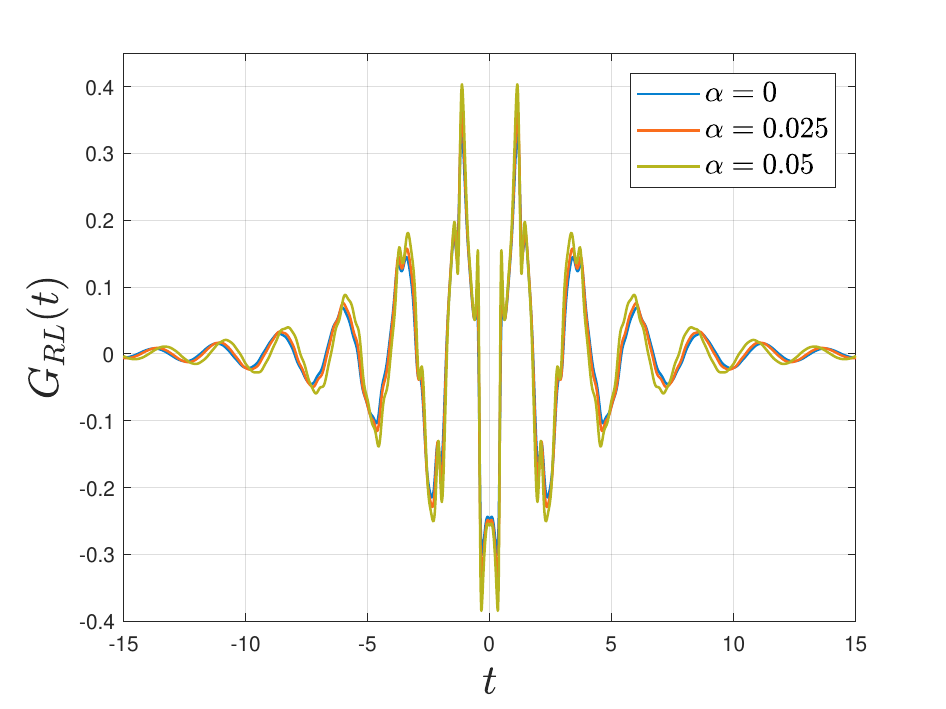}
        \caption{}
    \end{subfigure} 
    \caption{Various non-Hermtiain strengths of the off-diagonal Green's functions (a) $G_{LR}(t)$, (b) $G_{RL}(t)$ with $T=0.02, \mu=0.05$.}
\label{Fig.11}
\end{figure}

\begin{figure}[h]
\begin{subfigure}{0.45\linewidth}
        \centering        \includegraphics[width=3.2in]{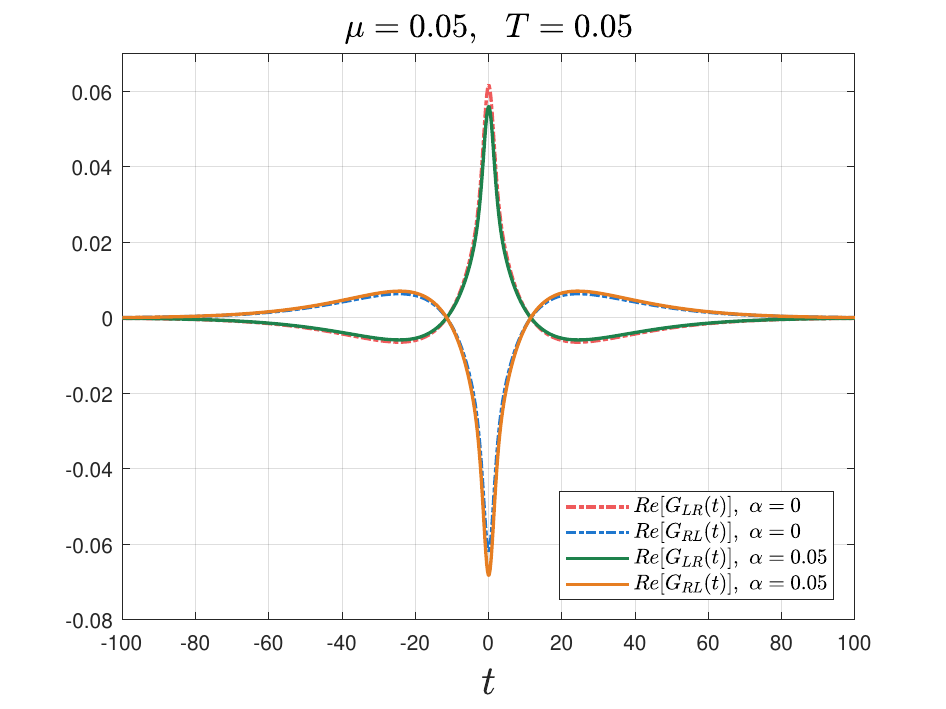}
        \caption{}
    \end{subfigure}
    \begin{subfigure}{0.45\linewidth}
        \centering
        \includegraphics[width=3.2in]{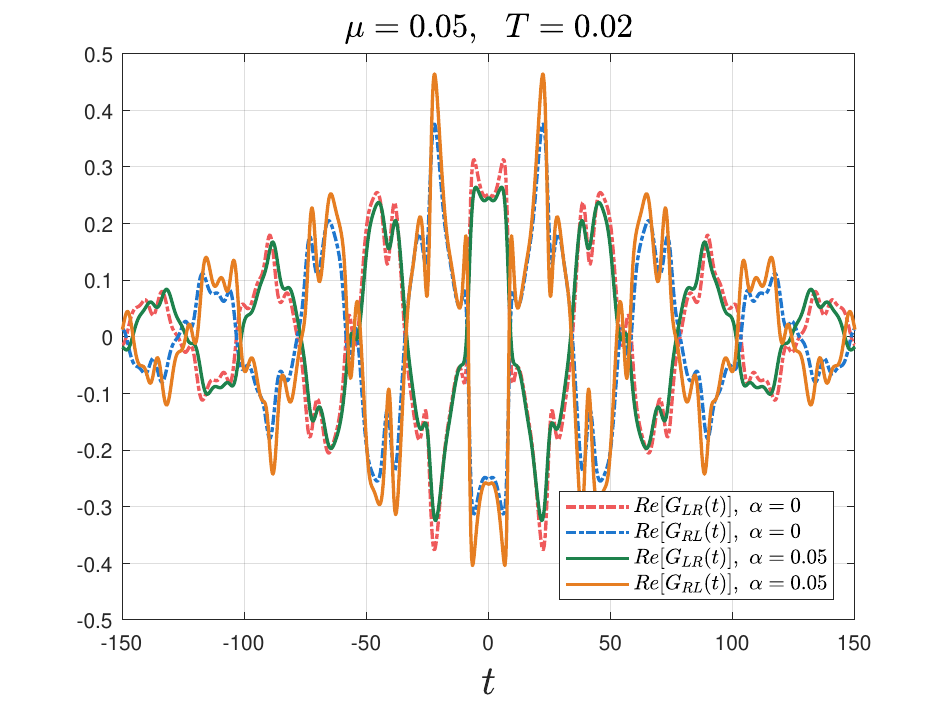}
        \caption{}
    \end{subfigure} 
    \caption{The real part of off-diagonal terms of Green’s functions Re[$G_{LR}(t)$] and Re[$G_{RL}(t)$] as functions of time for temperature (a) $T = 0.05$ (black hole phase) and (b) $T = 0.05$ (wormhole phase) with non-Hermitian parameter $\alpha = 0.05$.}
\label{Fig.12}
\end{figure}
\begin{figure}[h]
\begin{subfigure}{0.45\linewidth}
        \centering        \includegraphics[width=3.2in]{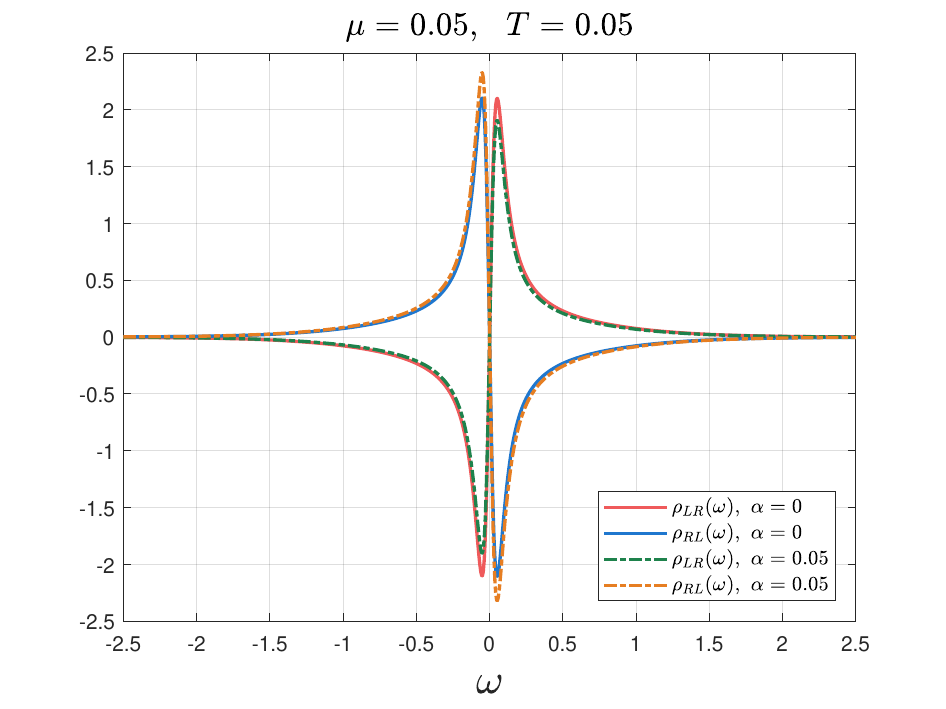}
        \caption{}
    \end{subfigure}
    \begin{subfigure}{0.45\linewidth}
        \centering
        \includegraphics[width=3.2in]{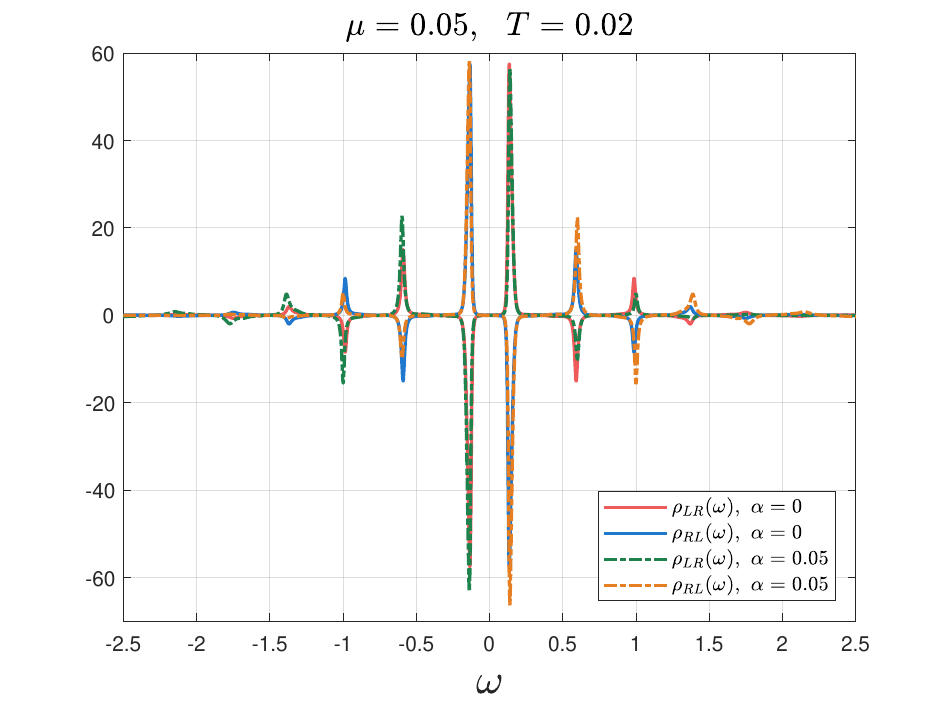}
        \caption{}
    \end{subfigure} 
    \caption{Spectral function $\rho_{LR}(\omega)$ and $\rho_{RL}(\omega)$ as functions of frequency, showing the impact of different parameters $\alpha,~T$ and $\mu$. (a) Spectral functions in the black hole phase at $\mu=0.05,~T=0.05$, with $\alpha=0$ and $\alpha=0.05$. (b) Spectral functions in the wormhole phase at $\mu=0.05,~T=0.02$, with $\alpha=0$ and $\alpha=0.05$.}
\label{Fig.13}
\end{figure}
\begin{figure}[h]
\begin{subfigure}{0.45\linewidth}
        \centering        \includegraphics[width=3.2in]{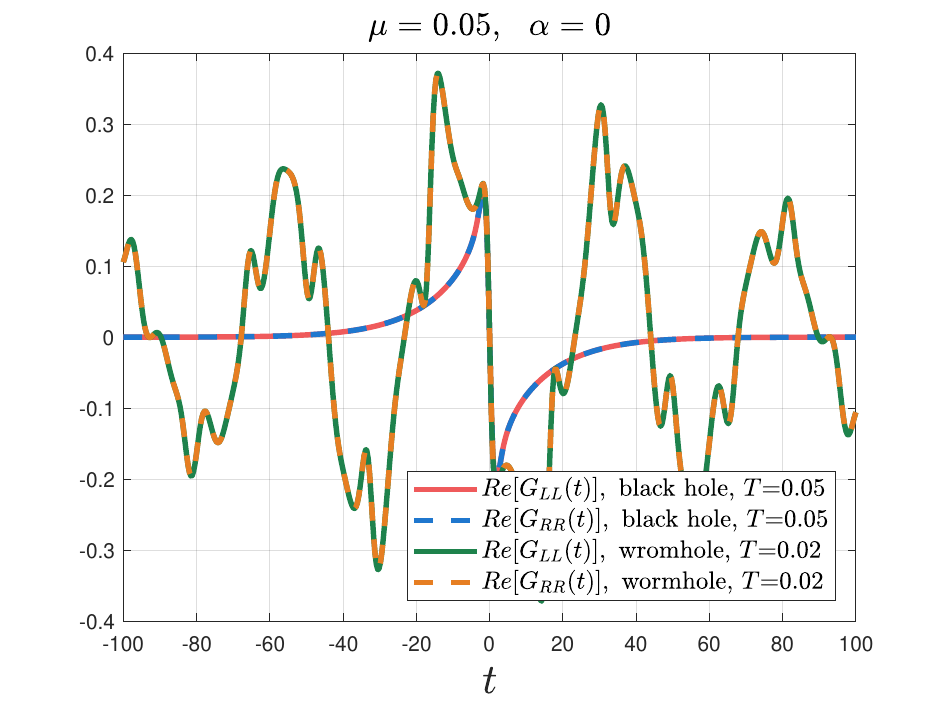}
        \caption{}
    \end{subfigure}
    \begin{subfigure}{0.45\linewidth}
        \centering
        \includegraphics[width=3.2in]{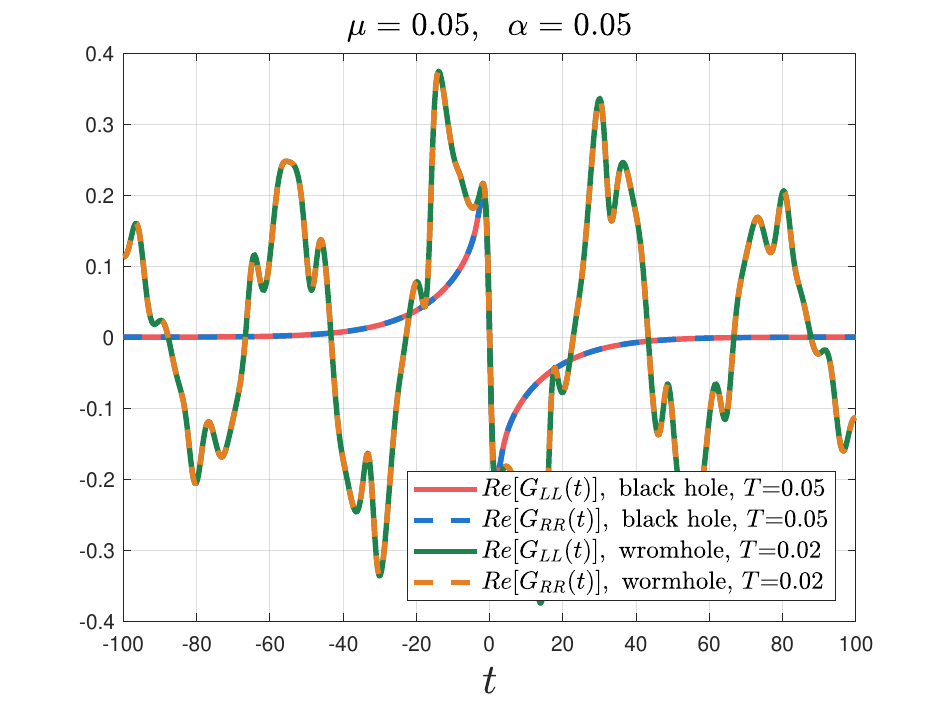}
        \caption{}
    \end{subfigure} 
    \caption{The diagonal Green’s functions $G_{LL}(t)$ and $G_{RR}(t)$ for temperature (a) $\alpha=0$ and (b) non-Hermitian parameter $\alpha=0.05$. Same as Fig.\ref{Diagnoal_Correlators}, the non-Hermitian parameter has a minor effect on the diagonal terms.}
\label{Fig.14}
\end{figure}

\begin{figure}[h]
\begin{subfigure}{0.45\linewidth}
        \centering        \includegraphics[width=3.2in]{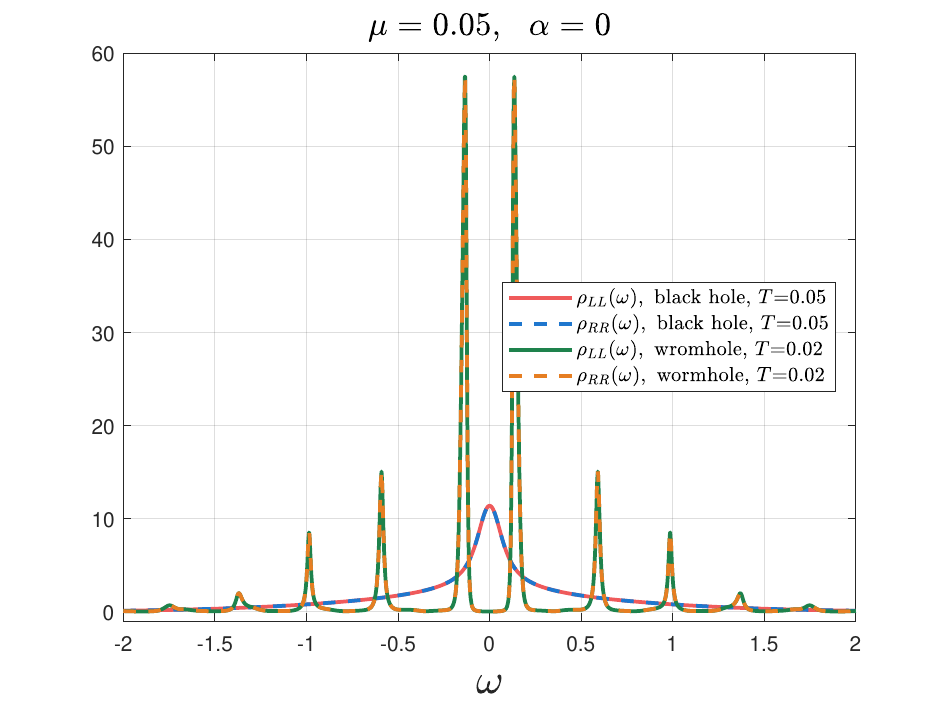}
        \caption{}
    \end{subfigure}
    \begin{subfigure}{0.45\linewidth}
        \centering
        \includegraphics[width=3.2in]{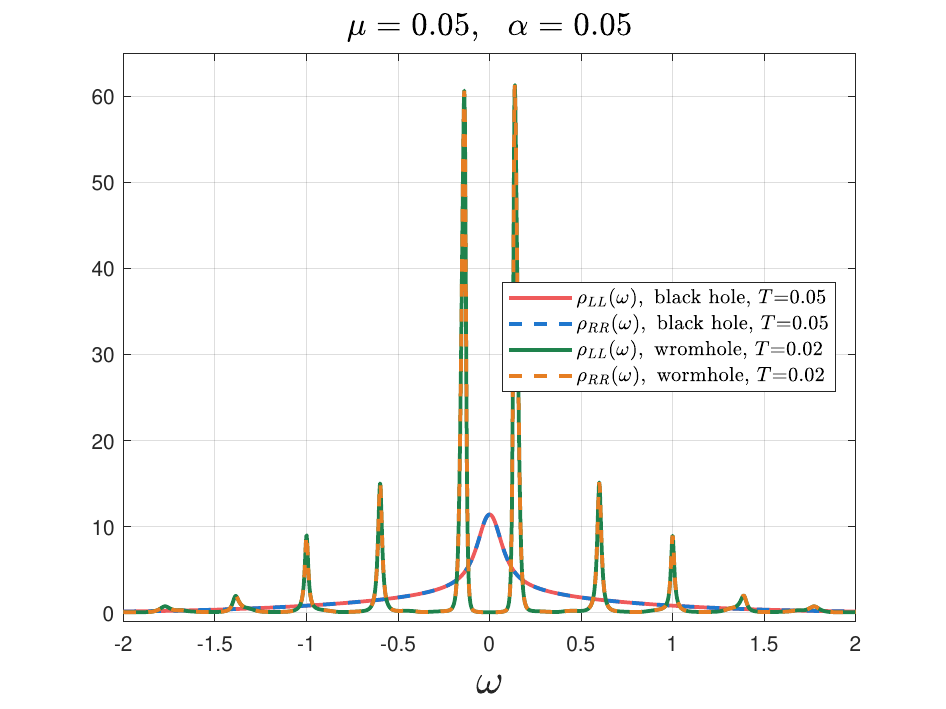}
        \caption{}
    \end{subfigure} 
    \caption{Diagonal spectral functions $\rho_{LL}(\omega)$ and $\rho_{RR}(\omega)$ for temperature $T = 0.05$ for black hole phase and $T=0.02$ for wormhole phase with (a) $\alpha=0$ and (b) $\alpha = 0.05$. Due to smaller $\mu$ and relatively not low temperature, unlike Fig.\ref{Diagnoal_spectrum_functions}, the spectral function in the black hole phase does not exhibit a concave structure at $\omega = 0$, closer to the behavior of non-interacting SYK model.}
\label{Fig.15}
\end{figure}

To further elucidate the spectral characteristics of the non-Hermitian model, we have carried out additional computations, the results of which are illustrated in Fig.~\ref{rho_small_mu}. This analysis reveals that the non-Hermitian model exhibits spectral function features akin to those detailed in Ref.~\cite{2003-03914}, particularly pertaining to low temperatures and small tunneling strengths, characterized by a more pronounced distribution of discrete peaks. As the coupling constant \(\mu\) decreases, more discrete peaks appear at larger \(\omega\) values, while the area ratio of the first peak becomes smaller. Specifically, the area ratio of the first peak is 60.3\% for \(\mu=0.05\) and 35.6\% for \(\mu=0.01\). The additional  calculation presented here allows us to underscore both the parallels and divergences in the spectral function characteristics between the Hermitian and non-Hermitian scenarios without disrupting the main narrative flow. Furthermore, Fig.~\ref{Fig.11} showcases the variation of non-diagonal Green functions across a range of 
$\alpha$ configurations. It becomes strikingly apparent that as 
$\alpha$ escalates, the magnitude of 
$G_{LR}$   wanes, juxtaposed with a corresponding surge in the amplitude of $G_{RL}$.

We also extend our analysis of the Green's functions and spectral functions to explore their behavior under different parameter settings. The additional figures presented here correspond to Figs.\ref{Fig.12} to \ref{Fig.15} but with different values of the non-Hermitian parameter $\alpha$ and temperature $T$.

\end{document}